\documentclass[10pt,aps,onecolumn,prd,superscriptaddress,nofootinbib,floatfix]{revtex4-2}
\usepackage{amsmath,amssymb,amsfonts}
\usepackage[utf8]{inputenc}
\usepackage[T1]{fontenc}
\usepackage{graphicx}
\usepackage{booktabs,array,colortbl,xcolor,longtable}
\usepackage{orcidlink}

\usepackage{hyperref}
\hypersetup{
  colorlinks=true,
  linkcolor=blue,
  citecolor=blue,
  urlcolor=blue
}

\makeatletter

\providecommand{\ifstrempty}[1]{%
  \if\relax\detokenize{#1}\relax
    \expandafter\@firstoftwo
  \else
    \expandafter\@secondoftwo
  \fi
}

\providecommand{\hrefbibentry}[2]{%
  \if\relax\detokenize{#1}\relax
    #2%
  \else
    \href{#1}{#2}%
  \fi
}

\makeatother

\newcolumntype{L}[1]{%
  >{\begin{minipage}[t]{#1}\raggedright\arraybackslash}%
  l%
  <{\end{minipage}}%
}

\begin{document}

\title{Geodesics and thermodynamics of a $\kappa$-deformed anti-de Sitter black hole surrounded by a quintessence field}

\author{Faizuddin Ahmed\orcidlink{0000-0003-2196-9622}}
\email{fahmed@rgu.ac ; faizuddin@associates.iucaa.in}
\affiliation{Department of Physics, The Assam Royal Global University, Guwahati, 781035, Assam, India}

\author{Ahmad Al-Badawi\orcidlink{0000-0002-3127-3453}}
\email{ahmadbadawi@ahu.edu.jo (Corresponding author)}
\affiliation{Department of Physics, Al-Hussein Bin Talal University, Ma'an 71111, Jordan}

\author{\.{I}zzet Sakall{\i}\orcidlink{0000-0001-7827-9476}}
\email{izzet.sakalli@emu.edu.tr}
\affiliation{Department of Physics, Eastern Mediterranean University, Famagusta 99628, North Cyprus via Mersin 10, Turkey\vspace{0.5cm}}

\author{Erdem Sucu\,\orcidlink{0009-0000-3619-1492}}
\email{erdemsc07@gmail.com}
\affiliation{Department of Physics, Eastern Mediterranean University, Famagusta 99628, North Cyprus via Mersin 10, Turkey\vspace{0.5cm}}
\begin{abstract}

We investigate the geodesic structure and thermodynamics of a $\kappa$-deformed Schwarzschild-anti-de Sitter black hole in a Kiselev quintessence field, where dynamics are governed by a mass-dependent lapse deformation. This term produces an inner Cauchy horizon without requiring charge or spin, yet it fails to resolve the central singularity; instead, the Kretschmann scalar diverges more strongly as $r^{-10}$. Because the deformation amplitude couples to the mass, the standard Bekenstein-Hawking area law violates the first law of thermodynamics. We resolve this by either deriving a modified entropy or rescaling the mass, both of which restore the first law and yield a consistent Smarr relation. Optically, an increasing deformation parameter shrinks both the photon sphere and the critical impact parameter. We also derive a closed-form Joule-Thomson coefficient, revealing that the $\kappa$-deformation triggers an inversion curve a feature absent in standard neutral SAdS. Notably, this occurs without generating a van der Waals critical point, demonstrating that Joule-Thomson inversion and van der Waals criticality can be cleanly decoupled. Finally, the deformation suppresses peak Hawking emission by a factor of roughly four and increases flux sparsity, an effect driven almost entirely by a drop in temperature rather than changes to the black hole shadow.

\end{abstract}

\maketitle

\section{Introduction}\label{isec1}

Quantum-corrected black hole solutions and the thermodynamics built on them have been examined closely over the past two decades, and the corrections turn out to carry phase transitions, criticality and Joule-Thomson expansion with them \cite{Snyder:1946qz,Doplicher:1994zv,Douglas:2001ba}. Most such solutions are constructed inside a specific approach to quantum gravity, usually loop quantum gravity or a noncommutative spacetime geometry. Noncommutative geometry captures the quantum structure of spacetime through a fundamental length scale that enters the field equations as a smeared matter distribution, an idea with roots in string theory~\cite{Seiberg:1999vs} \cite{Snyder:1946qz,Doplicher:1994zv}. The minimal length of Moyal spacetime, which is noncommutative of canonical type \cite{Douglas:2001ba}, replaces the classical curvature singularity with a regular de Sitter core \cite{Nicolini:2005vd}. The minimal length of $\kappa$-deformed spacetime, which is noncommutative of Lie-algebraic type \cite{Arzano:2021hpg}, removes the big-bang singularity through a bounce \cite{Rajagopal:2025qyp}, in the same spirit as the quantum Oppenheimer-Snyder models of Refs.~\cite{Lewandowski:2022zce}. Both parameters reshape the phase transition, the critical ratio and the inversion temperature of standard black hole thermodynamics \cite{Banerjee:2008gc,Modesto:2010rv,Liang:2017rng,Filho:2022zdh,Filho:2024zxx,Panja:2026hja}.

The Moyal parameter converts the Hawking-Page transition from a plain first-order transition into a van der Waals phase diagram with a critical point, past which the transition softens into a crossover \cite{Nicolini:2011dp}. When the noncommutative parameter and its conjugate are promoted to thermodynamic variables in the extended phase space, both the Moyal and the $\kappa$-deformed parameters generate phase transitions and criticality in uncharged Schwarzschild-anti-de Sitter (AdS) black holes \cite{Wang:2024jlj,Tan:2024jkj,Wang:2025ycl,Wang:2025alf,Kumara:2026uwi}. The loop-quantum-gravity parameter of the quantum-corrected Schwarzschild-AdS solution does the same \cite{Wang:2024jtp}.

The characteristic noncommutative \cite{Wang:2024jlj,Tan:2024jkj,Wang:2025ycl,Wang:2025alf,MoraisGraca:2021ife} and loop-quantum-gravity \cite{Wang:2024jtp} corrections to Schwarzschild-AdS introduce a Joule-Thomson effect in black holes that carry neither charge nor angular momentum, and the ratio of the minimum inversion temperature to the critical temperature comes out independent of the minimal length scale. Whether this is a universal property of quantum-gravity-corrected black holes, or something peculiar to the Moyal and loop-quantum-gravity families, is not settled. Answering it requires a throttling analysis of the $\kappa$-deformed Schwarzschild-AdS solution, and that is one of the two aims of this work. We take the solution constructed in Ref.~\cite{Kumara:2026uwi} and immerse it in a Kiselev quintessence field \cite{Kiselev2003}, so that the deformation is tested against a dark-energy background rather than in vacuum. Geometries surrounded by a quintessence field exhibit modified horizon structure and thermodynamic properties, with the quintessence parameter generally altering the locations and number of horizons as well as the black-hole causal structure \cite{ALBADAWI2026117489,FA10,FA5,FA2}.

The second aim concerns the geodesic structure, which is where a deformation of this kind becomes observationally exposed. Photon spheres, critical impact parameters and shadow radii of quintessence-dressed black holes have been mapped in a number of settings \cite{Uniyal:2024hay,Nozari:2024ncq,Sood:2024sha,Sakalli:2026abg}, as have the innermost stable circular orbits and epicyclic frequencies that set quasiperiodic oscillation models \cite{Jha:2026kro,Sakalli:2025geo}. Quantum-corrected geometries have been examined along the same lines \cite{Ali:2022geo,Sakalli:2020qui,Filho:2024zxx,Sakalli:2025eqg,Sakalli:2025gmn}.  What is missing is the corresponding treatment of the $\kappa$-deformed geometry, where the deformation amplitude scales as $M^{3}$ rather than as an independent hair, and where that scaling turns out to have consequences well beyond the optics.

That scaling is the thread running through this paper. Writing the deformation as $\hat a M^{3}/2\pi r^{3}$ with $\hat a=a/M$ dimensionless means the coefficient of $r^{-3}$ is not a free constant of the solution but a quantity that changes as the black hole radiates. Three things follow. First, the horizon equation becomes a cubic in $M$ rather than a linear one, and the branch connected to the undeformed mass survives only inside a bounded region of parameter space. Second, the area law and the first law become mutually inconsistent, so that one of the two has to be corrected; we correct the entropy and verify the outcome numerically. Third, the Joule-Thomson coefficient acquires a compact closed form whose zero locus is a one-derivative shift of the photon-sphere condition, which ties the throttling behaviour directly to the null geodesic structure of the same geometry.

The paper is organised as follows. Section~\ref{isec2} sets up the geometry, lists the limits in which it reduces to known solutions, computes the curvature invariants and establishes the admissibility bound on the deformation. Section~\ref{isec3} treats null geodesics: the photon sphere, the critical impact parameter, the shadow seen by a static observer at finite distance, the Lyapunov exponent and the eikonal ringdown estimate. Section~\ref{isec4} turns to massive particles, circular orbits, the innermost stable circular orbit and the epicyclic frequencies. Section~\ref{isec5} builds the thermodynamics, derives the corrected entropy and checks the first law. Section~\ref{isec6} obtains the Joule-Thomson coefficient, maps the inversion curves and contrasts the outcome with the Moyal and loop-quantum-gravity cases. Section~\ref{isec7} treats the sparsity of the Hawking flux and the energy emission rate. Section~\ref{isec8} collects the results. We use geometric units $G=c=1$ and signature $(-,+,+,+)$, and set $M=1$ in all numerical work.

\section{The $\kappa$-deformed AdS black hole in a quintessence field}\label{isec2}

This section fixes the geometry we work with and establishes the properties that constrain everything downstream. We begin from the solution of Ref.~\cite{Kumara:2026uwi}, add the Kiselev quintessence tail, and then work through the limits, the curvature invariants, the horizon structure and the range of the deformation parameter over which the solution exists at all.   The last of these turns out to be restrictive in a way that has not been noted before, and it fixes the parameter windows used in every figure of this paper.

The minimal-length corrections of noncommutative spacetime structures \cite{Nicolini:2005vd} and of T-duality \cite{Nicolini:2019irw} deform a point-like matter distribution and feed quantum corrections into the energy-momentum tensor, which then propagate to the metric through the Einstein equations. The $\kappa$-deformed Schwarzschild-AdS solution was built in Ref.~\cite{Kumara:2026uwi} along exactly this route, by solving the field equations with the corrected energy-momentum tensor that the $\kappa$-deformation parameter generates. Its energy density and pressure components follow from inserting the $\kappa$-deformed Newtonian potential into the Poisson equation, consistent with the conservation law. Adding the Kiselev tail \cite{Kiselev2003} gives the line element
\begin{equation}
ds^{2}=-f(r)\,dt^{2}+f(r)^{-1}dr^{2}+r^{2}d\Omega^{2},
\label{eq:metric}
\end{equation}
with
\begin{equation}
f(r)=1-\frac{2M}{r}+\frac{\hat a M^{3}}{2\pi r^{3}}-\frac{\Lambda r^{2}}{3}-\frac{N}{r^{3w+1}},
\label{eq:lapse}
\end{equation}
where $\hat a=a/M$ is the dimensionless $\kappa$-deformation parameter in natural units \cite{Kumara:2026uwi}, and $(N,w)$ are the quintessence parameters \cite{Kiselev2003}. We take $\Lambda<0$ throughout and write $\Lambda=-3/\ell^{2}$, with $P=-\Lambda/8\pi=3/8\pi\ell^{2}$ the thermodynamic pressure. Quintessence proper occupies $-1<w<-1/3$, and we use $w=-2/3$ in the numerics, for which the tail is linear, $-Nr$.

The lapse in Eq.~\eqref{eq:lapse} contains four independent structures, and switching them off one at a time recovers a chain of known geometries. Table~\ref{tab:limits} lists them together with the quantity that is easiest to check in each case, and we have verified each entry numerically. Two entries deserve comment. At $w=-1$ the tail becomes $-Nr^{2}$, which is indistinguishable from a shift of the cosmological constant, so the solution collapses onto the $\kappa$-deformed Schwarzschild-AdS geometry with $\Lambda_{\rm eff}=\Lambda+3N$. At $w=-1/3$ the tail is the constant $-N$, so with $\Lambda=0$ the metric is not asymptotically flat but carries a solid angle deficit, $f(\infty)=1-N$. The shadow radius then acquires a factor $\sqrt{f(\infty)}$ that is easy to drop and that reverses how the shadow scales with $N$; we return to this in Sec.~\ref{isec3}.

\begin{longtable}{L{0.2164\textwidth} L{0.2705\textwidth} L{0.3246\textwidth}}
  \hline\hline
\cellcolor{brown!30}\textbf{Parameter choice} & \cellcolor{brown!30}\textbf{Geometry recovered} & \cellcolor{brown!30}\textbf{Verified quantity} \\
  \hline
  \endfirsthead
  \endhead
  \hline
  \endfoot
  \hline\hline
  \caption{Limiting cases of the lapse in Eq.~\eqref{eq:lapse}. Each row was checked numerically against the closed-form result quoted in the third column; all forty checks agree to better than one part in $10^{6}$.}
  \label{tab:limits}\\
  \endlastfoot
$\hat a=N=\Lambda=0$ & Schwarzschild & $r_{h}=2M$, $r_{\rm ph}=3M$, $b_{c}=3\sqrt3 M$, $r_{\rm ISCO}=6M$, $T=1/8\pi M$, $K=48M^{2}/r^{6}$ \\
$\hat a=N=0$ & Schwarzschild-AdS & $R=4\Lambda$, $T=(1-\Lambda r_{h}^{2})/4\pi r_{h}$, $r_{\rm ph}=3M$ exactly \\
$\hat a=0$ & Kiselev-AdS & $\rho_{q}=-3wN/8\pi r^{3w+3}$, $\langle p\rangle/\rho_{q}=w$ \\
$N=0$ & $\kappa$-deformed Schwarzschild-AdS \cite{Kumara:2026uwi} & extremal at $\hat a=64\pi/27$, $r_{\rm ext}=4M/3$ \\
$w=-1$ & $\kappa$-deformed Schwarzschild-AdS & $\Lambda_{\rm eff}=\Lambda+3N$ \\
$w=-1/3$, $\Lambda=0$ & solid angle deficit & $f(\infty)=1-N$, $R_{\rm sh}=b_{c}\sqrt{1-N}$, $r_{\rm ph}=3M/(1-N)$ \\
$w=1/3$, $N=-Q^{2}$ & Reissner-Nordstr\"om & $f=1-2M/r+Q^{2}/r^{2}$ \\
$M=\hat a=N=0$ & pure AdS & $K=8\Lambda^{2}/3$ \\
\end{longtable}

\subsection{Curvature invariants and the fate of the singularity}\label{isec2a}

For a lapse of the form Eq.~\eqref{eq:metric} the Ricci scalar and the Kretschmann invariant reduce to
\begin{equation}
R=-f''-\frac{4f'}{r}-\frac{2f}{r^{2}}+\frac{2}{r^{2}},\qquad
K=f''^{2}+\frac{4f'^{2}}{r^{2}}+\frac{4(f-1)^{2}}{r^{4}}.
\label{eq:invariants}
\end{equation}
Substituting Eq.~\eqref{eq:lapse} gives
\begin{equation}
R=4\Lambda-\frac{\hat a M^{3}}{\pi r^{5}}+\frac{3wN(3w-1)}{r^{3w+3}},
\label{eq:ricci}
\end{equation}
which is finite everywhere except at the origin and which tends to $4\Lambda$ as $r\to\infty$ for $w>-1$, confirming the AdS asymptotics. The Kretschmann scalar is longer, but its behaviour near the origin is governed by the deformation alone,
\begin{equation}
K\;\longrightarrow\;\frac{46\,\hat a^{2}M^{6}}{\pi^{2}r^{10}}\qquad (r\to0),
\label{eq:kret_origin}
\end{equation}
a limit we obtained symbolically and checked at $w=-1$, $w=-2/3$ and $w=-1/3$ with identical results. Equation~\eqref{eq:kret_origin} settles a question that the bounce result of Ref.~\cite{Rajagopal:2025qyp} might otherwise leave open. The $\kappa$-deformation does not regularise this black hole. It makes the central singularity worse, replacing the Schwarzschild divergence $48M^{2}/r^{6}$ with an $r^{-10}$ divergence whose coefficient grows quadratically in $\hat a$. This is the behaviour expected on general grounds for a power-law addition to the lapse: a term $\gamma r^{-p}$ contributes to $K$ at order $r^{-2p-4}$, and no choice of $\gamma$ cancels it. A regular core requires $f=1-2m(r)/r$ with $m(r)\sim r^{3}$ near the origin, which the $r^{-3}$ deformation does not provide. Figure~\ref{fig:kretschmann} shows the crossover between the two regimes.

\subsection{Effective source and the null energy condition}\label{isec2b}

Reading Eq.~\eqref{eq:lapse} as a solution of the Einstein equations with an effective source, the energy density and the pressures are
\begin{equation}
\rho=-p_{r}=\frac{1-f-rf'}{8\pi r^{2}},\qquad
p_{t}=\frac{1}{16\pi}\Big(f''+\frac{2f'}{r}\Big),
\label{eq:source}
\end{equation}
so that
\begin{equation}
\rho=\frac{\Lambda}{8\pi}+\frac{\hat a M^{3}}{8\pi^{2}r^{5}}-\frac{3wN}{8\pi r^{3w+3}}.
\label{eq:rho}
\end{equation}
The quintessence sector reproduces Kiselev's equation of state exactly: with $M=\hat a=\Lambda=0$ the isotropic average $\langle p\rangle=(p_{r}+2p_{t})/3$ satisfies $\langle p\rangle=w\rho$, which we confirmed numerically to machine precision. The radial null energy condition is saturated identically, $\rho+p_{r}=0$, as it must be for any metric with $g_{tt}g_{rr}=-1$.

\begin{figure}[ht!]
  \centering
  \includegraphics[width=0.78\textwidth]{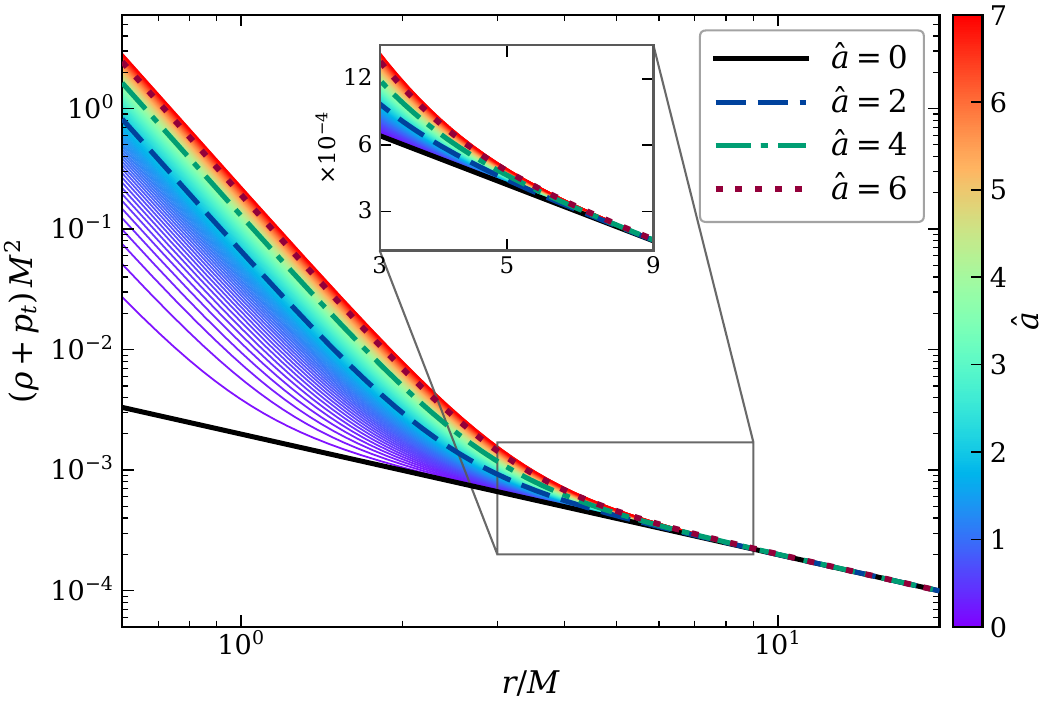}
  \caption{Tangential null energy combination of Eq.~\eqref{eq:rho_pt} on logarithmic axes, for $\Lambda=-0.03$, $N=0.05$, $w=-2/3$, with the ribbon sweeping $\hat a$ over $[0,7]$. The combination is positive throughout. The inset resolves $r/M\in[3,9]$, where the curves merge onto the quintessence envelope $N/8\pi r$.}
  \label{fig:nec}
\end{figure}

The tangential combination is the informative one, and it collapses to something simpler than Eq.~\eqref{eq:rho} would suggest. Using $\rho+p_{t}=[2(1-f)/r^{2}+f'']/16\pi$ and inserting Eq.~\eqref{eq:lapse},
\begin{equation}
\rho+p_{t}=\frac{1}{16\pi^{2}}\left[\frac{5\hat aM^{3}}{r^{5}}-\frac{9\pi Nw(w+1)}{r^{3w+3}}\right].
\label{eq:rho_pt}
\end{equation}
Both the mass and the cosmological constant cancel out of this expression, which is worth pausing on: the Schwarzschild and anti-de Sitter parts of the geometry contribute nothing to the tangential null energy condition, and only the deformation and the quintessence do. For $\hat a\ge0$ the first term is positive, and for quintessence proper, $-1<w<-1/3$, the factor $-w(w+1)$ is positive as well, so
\begin{equation}
\rho+p_{t}\ge0
\end{equation}
everywhere, with equality only in the Schwarzschild-anti-de Sitter limit $\hat a=N=0$, where the combination vanishes identically. The tangential null energy condition is therefore satisfied rather than violated, and the deformation strengthens rather than weakens it, in contrast with the minimal-length constructions of Refs.~\cite{Hamil:2022barrow,Bolen:2008gup} where the corrected source does not retain a definite sign. Figure~\ref{fig:nec} shows the two regimes on logarithmic axes. Inside a few $M$ the $r^{-5}$ deformation term dominates and the combination rises steeply with $\hat a$; outside, the quintessence tail takes over and every curve collapses onto the common $r^{-(3w+3)}$ envelope, which for $w=-2/3$ is $N/8\pi r$ and carries no $\hat a$ dependence at all.

\subsection{Horizons and the admissibility bound}\label{isec2c}

\begin{figure}[ht!]
  \centering
  \includegraphics[width=0.78\textwidth]{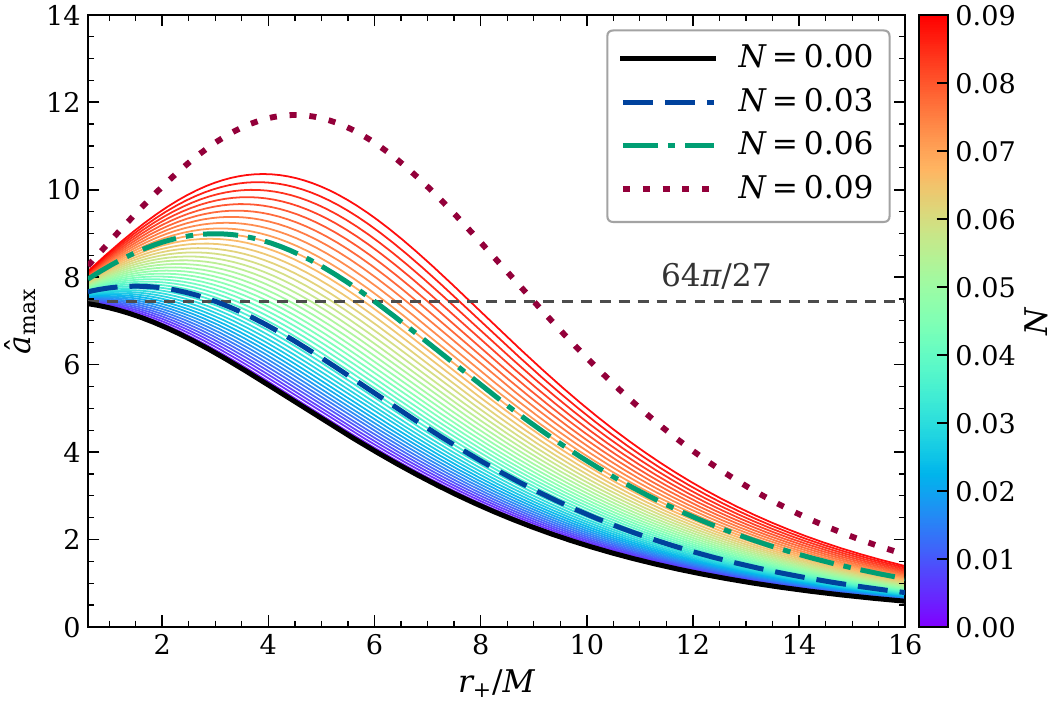}
  \caption{Admissibility bound $\hat a_{\max}=64\pi/27G^{2}$ of Eq.~\eqref{eq:amax} against horizon radius, for $\Lambda=-0.03$, $w=-2/3$, with the ribbon sweeping $N$ over $[0,0.09]$. The dashed horizontal line marks the Schwarzschild-sector value $64\pi/27=7.4467$. The bound peaks near $r_{+}\simeq3M$ before the anti-de Sitter term drives it down.}
  \label{fig:admissible}
\end{figure}

Setting $f(r_{+})=0$ in Eq.~\eqref{eq:lapse} and clearing denominators produces a cubic in the mass,
\begin{equation}
M^{3}-\frac{4\pi r_{+}^{2}}{\hat a}\,M+\frac{2\pi r_{+}^{3}}{\hat a}\,G(r_{+})=0,\qquad
G(r_{+})\equiv1-\frac{\Lambda r_{+}^{2}}{3}-\frac{N}{r_{+}^{3w+1}},
\label{eq:cubic}
\end{equation}
which is the first structural consequence of tying the deformation amplitude to the mass. For $\hat a\to0$ the relevant root approaches the familiar $M_{0}=r_{+}G(r_{+})/2$. For $\hat a>0$ the depressed cubic has three real roots whenever its discriminant is positive, and that requirement reads
\begin{equation}
\hat a<\hat a_{\max}(r_{+})=\frac{64\pi}{27\,G(r_{+})^{2}}.
\label{eq:amax}
\end{equation}
In the Schwarzschild sector $G=1$ and the bound is the pure number $64\pi/27=7.4467$, which we independently located as the value at which the inner and outer horizons merge at $r_{\rm ext}=4M/3$. Equation~\eqref{eq:amax} is more than a technical restriction. Because $G$ grows with $r_{+}$ once the AdS term dominates, $\hat a_{\max}$ falls, and a black hole of given deformation strength cannot be arbitrarily large. Figure~\ref{fig:admissible} maps the bound, and it is the reason every figure in this paper stops at $\hat a\simeq7$~\cite{Ahmed:2026ncq,Sood:2026cos}.

Of the three roots of Eq.~\eqref{eq:cubic}, one is negative and two are positive. The smaller positive root is the branch on which $r_{+}$ is the event horizon, since it is the one satisfying $f'(r_{+})>0$; the larger positive root places $r_{+}$ at an inner horizon instead. Writing the trigonometric solution of the depressed cubic, the physical branch is
\begin{equation}
M(r_{+})=\frac{4r_{+}}{\sqrt{3}}\sqrt{\frac{\pi}{\hat a}}\;
\cos\!\left[\frac{1}{3}\arccos\!\left(-\frac{3G(r_{+})}{4}\sqrt{\frac{3\hat a}{\pi}}\,\frac{1}{2}\right)-\frac{2\pi}{3}\right],
\label{eq:mass_branch}
\end{equation}
which we use throughout the numerical work and which reproduces the exact root of Eq.~\eqref{eq:cubic} to ten decimal places.

\begin{figure}[ht!]
  \centering
  \includegraphics[width=0.8\textwidth]{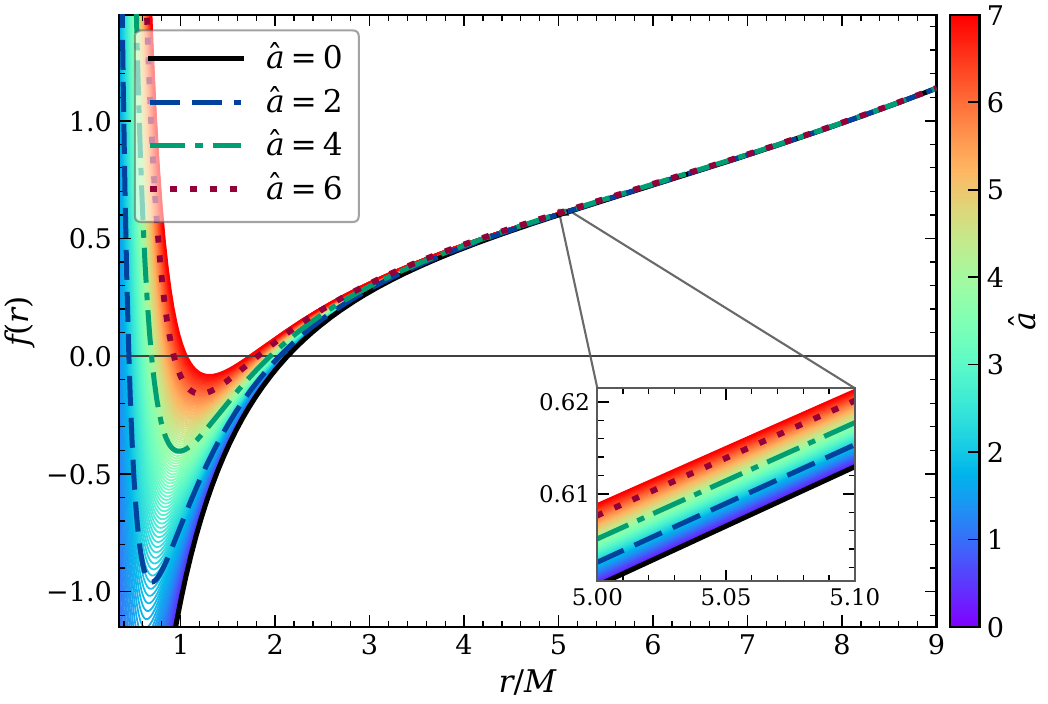}
  \caption{Lapse $f(r)$ of Eq.~\eqref{eq:lapse} for $\Lambda=-0.03$, $N=0.05$, $w=-2/3$. The ribbon sweeps $\hat a$ continuously over $[0,7]$ under the colour bar; the four labelled curves are drawn on top in distinct line styles. The inset magnifies $r/M\in[5.00,5.10]$, where the four curves are separated by less than $8\times10^{-3}$.}
  \label{fig:lapse}
\end{figure}

\begin{figure}[ht!]
  \centering
  \includegraphics[width=0.8\textwidth]{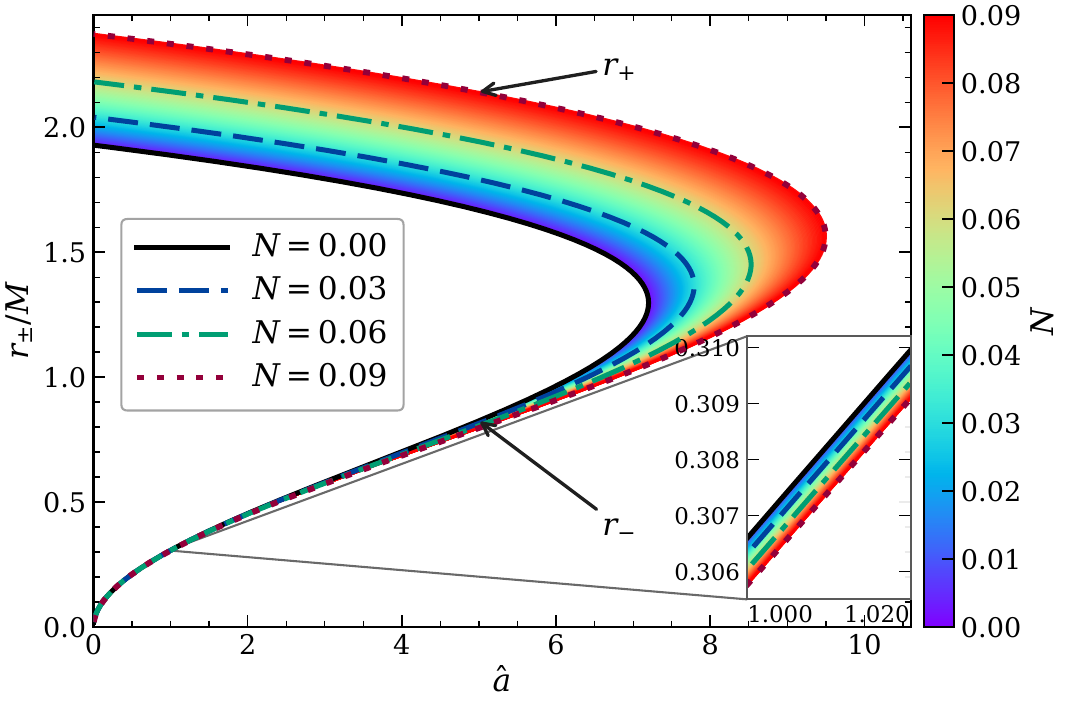}
  \caption{Inner and outer horizon radii, arrowed, against the deformation parameter for $\Lambda=-0.03$, $w=-2/3$. Each curve is traced by inverting $f(r_{+})=0$ for $\hat a$, so the two branches form one continuous fold; the ribbon sweeps the quintessence normalisation $N$ over $[0,0.09]$ and the labelled curves give four fixed values. The fold tip, where the horizons merge, moves from $\hat a=7.200$ at $N=0$ to $\hat a=9.497$ at $N=0.09$. The inset magnifies $\hat a\in[1.000,1.020]$ on the inner branch, where the four curves are separated by less than $4\times10^{-3}M$.}
  \label{fig:horizons}
\end{figure}

The horizon structure itself is shown in Figs.~\ref{fig:lapse} and~\ref{fig:horizons}. At $\hat a=0$ the geometry has a single horizon. Any $\hat a>0$ opens an inner Cauchy horizon, because the deformation term is positive and drives $f\to+\infty$ as $r\to0$, so $f$ must cross zero twice. This happens without charge and without rotation, which is worth stating plainly: the $\kappa$-deformation manufactures a Cauchy horizon out of the vacuum sector alone. As $\hat a$ climbs toward $\hat a_{\max}$ the two horizons approach each other and merge at extremality. Table~\ref{tab:horizons} records the numbers for the baseline configuration $\Lambda=-0.03$, $N=0.05$, $w=-2/3$: the outer horizon contracts from $2.1302M$ at $\hat a=0$ to $1.7209M$ at $\hat a=7$, while the inner horizon expands from zero to $1.0679M$, so the region between them shrinks by better than a factor of two over the admissible range.

\begin{figure}[ht!]
  \centering
  \includegraphics[width=0.78\textwidth]{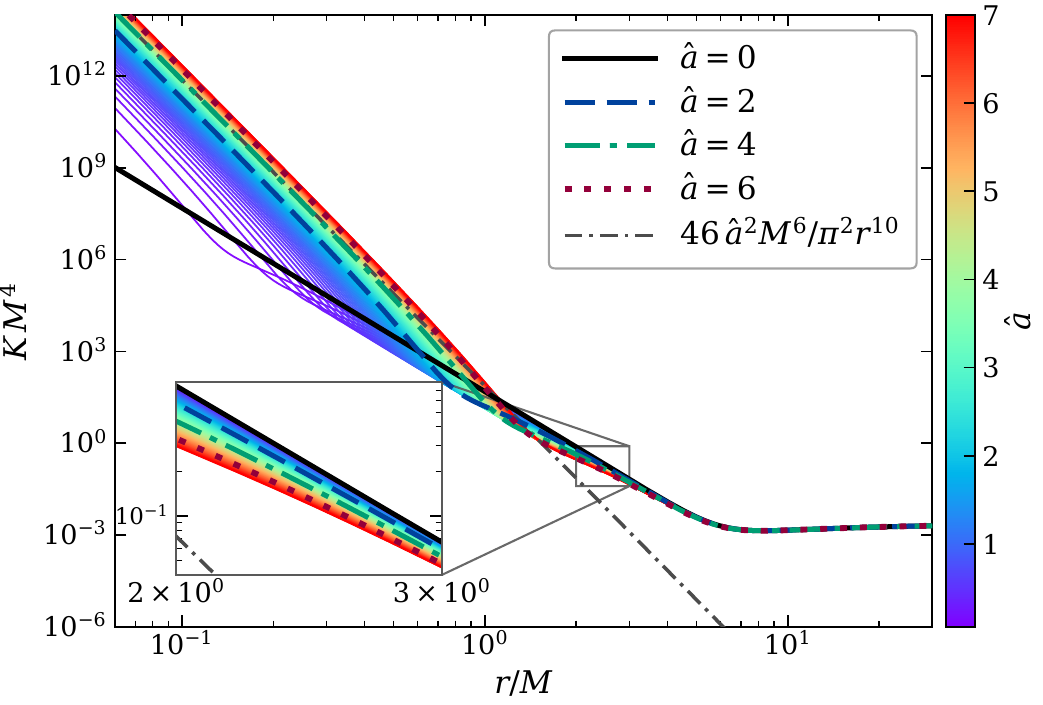}
  \caption{Kretschmann invariant on logarithmic axes for $\Lambda=-0.03$, $N=0.05$, $w=-2/3$, with the ribbon sweeping $\hat a$ over $(0,7]$. The grey dash-dotted reference line is the small-$r$ asymptote of Eq.~\eqref{eq:kret_origin} evaluated at $\hat a=4$. The inset covers $r/M\in[2.0,3.0]$, where the deformation and Schwarzschild contributions cross over.}
  \label{fig:kretschmann}
\end{figure}

\begin{longtable}{L{0.0802\textwidth} L{0.1123\textwidth} L{0.1123\textwidth} L{0.1284\textwidth} L{0.1123\textwidth} L{0.1123\textwidth}}
  \hline\hline
\cellcolor{brown!30}$\hat a$ & \cellcolor{brown!30}$r_{-}/M$ & \cellcolor{brown!30}$r_{+}/M$ & \cellcolor{brown!30}$(r_{+}-r_{-})/M$ & \cellcolor{brown!30}$r_{\rm ph}/M$ & \cellcolor{brown!30}$\hat a_{\max}$ \\
  \hline
  \endfirsthead
  \endhead
  \hline
  \endfoot
  \hline\hline
  \caption{Horizon structure for $M=1$, $\Lambda=-0.03$, $N=0.05$, $w=-2/3$. The last column evaluates the bound of Eq.~\eqref{eq:amax} at the corresponding outer horizon. Roots were located by bisection to a tolerance of $10^{-13}$.}
  \label{tab:horizons}\\
  \endlastfoot
0.0 & $-$ & 2.13023 & $-$ & 3.26680 & 8.4481 \\
1.0 & 0.30613 & 2.09076 & 1.78463 & 3.22102 & 8.4425 \\
2.0 & 0.45212 & 2.04791 & 1.59579 & 3.17257 & 8.4359 \\
3.0 & 0.57611 & 2.00079 & 1.42468 & 3.12096 & 8.4279 \\
4.0 & 0.69228 & 1.94807 & 1.25579 & 3.06559 & 8.4179 \\
5.0 & 0.80777 & 1.88753 & 1.07976 & 3.00562 & 8.4053 \\
6.0 & 0.92923 & 1.81507 & 0.88584 & 2.93984 & 8.3886 \\
7.0 & 1.06787 & 1.72094 & 0.65307 & 2.86643 & 8.3641 \\
\end{longtable}

\section{Null geodesics, photon spheres and the shadow}\label{isec3}

We now ask how the deformation shows up in the propagation of light, which is the channel through which a correction of this kind would first become visible. Staticity and spherical symmetry give the conserved energy $E=f\dot t$ and angular momentum $L=r^{2}\dot\phi$, and equatorial null geodesics obey $\dot r^{2}=E^{2}-V^{\rm null}_{\rm eff}$ with
\begin{equation}
V^{\rm null}_{\rm eff}(r)=\frac{L^{2}f(r)}{r^{2}}.
\label{eq:vnull}
\end{equation}
Circular photon orbits sit at the extrema of $f/r^{2}$, so the photon sphere solves
\begin{equation}
r f'(r)=2f(r).
\label{eq:phsphere}
\end{equation}
Inserting Eq.~\eqref{eq:lapse} and simplifying, the cosmological term cancels identically and the condition reduces to
\begin{equation}
-2+\frac{6M}{r}-\frac{5\hat a M^{3}}{2\pi r^{3}}+N(3w+3)\,r^{-(3w+1)}=0.
\label{eq:phsphere_explicit}
\end{equation}
The cancellation of $\Lambda$ is exact and worth recording: the photon sphere of Schwarzschild-AdS sits at $r_{\rm ph}=3M$ for every value of the cosmological constant, a result our numerics reproduce to twelve digits. The deformation and the quintessence tail, by contrast, both move it. The $\hat a$ term enters with a negative sign and pulls the photon sphere inward; the quintessence term at $w=-2/3$ contributes $+Nr$ and pushes it outward, and it also creates a second root. Solving Eq.~\eqref{eq:phsphere_explicit} for $w=-2/3$ at large radius gives $Nr\simeq2$, so there is an outer unstable photon orbit near $r\simeq2/N$, which for $N=0.05$ sits at $r\simeq40M$. Circular timelike orbits exist only between the two, a point we use in Sec.~\ref{isec4}.

The critical impact parameter follows from $b_{c}=r_{\rm ph}/\sqrt{f(r_{\rm ph})}$. Converting it into an observable requires care in an AdS background. For an asymptotically flat spacetime one writes $R_{\rm sh}=b_{c}\sqrt{f(\infty)}$, and this is where the $w=-1/3$ deficit noted in Sec.~\ref{isec2} matters: there $f(\infty)=1-N$, and omitting the square root reverses the direction in which the shadow responds to $N$. With $\Lambda<0$ the situation is different again, since $f$ diverges and no asymptotic observer exists. We therefore quote the angular radius measured by a static observer at finite areal radius $r_{o}$,
\begin{equation}
\sin^{2}\alpha_{\rm sh}=\frac{b_{c}^{2}\,f(r_{o})}{r_{o}^{2}},
\label{eq:shadow_angle}
\end{equation}
and state $r_{o}$ alongside every number. Table~\ref{tab:null} uses $r_{o}=50M$.

\begin{figure}[ht!]
  \centering
  \includegraphics[width=0.72\textwidth]{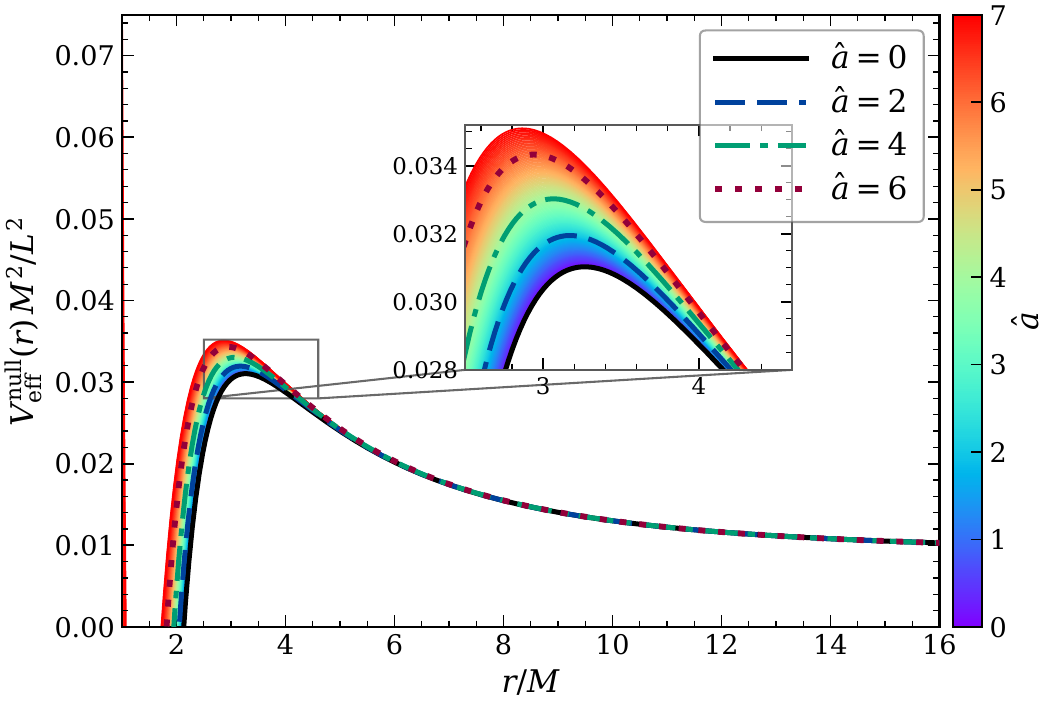}
  \caption{Null effective potential of Eq.~\eqref{eq:vnull} in units of $L^{2}/M^{2}$, for $\Lambda=-0.03$, $N=0.05$, $w=-2/3$, with the ribbon sweeping $\hat a$ over $[0,7]$. The inset spans $r/M\in[2.5,4.6]$ and the full height of the barrier maximum.}
  \label{fig:vnull}
\end{figure}

\begin{figure}[ht!]
  \centering
  \includegraphics[width=0.78\textwidth]{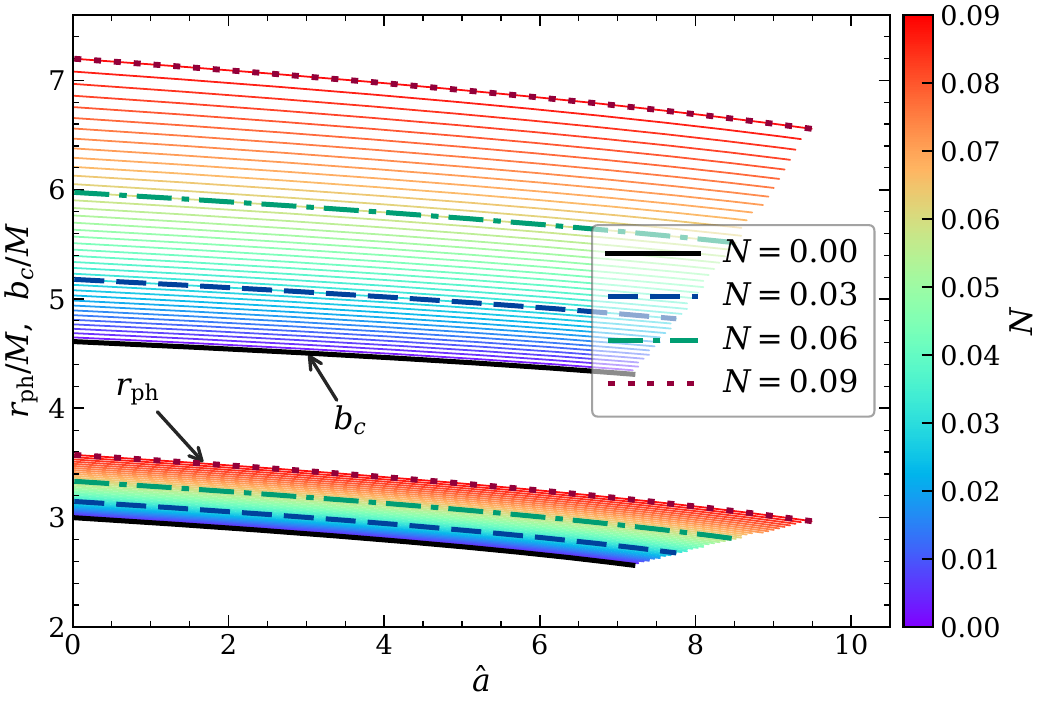}
  \caption{Photon-sphere radius (lower band) and critical impact parameter (upper band), both arrowed, against $\hat a$ on a common axis, for $\Lambda=-0.03$, $w=-2/3$, with the ribbon sweeping $N$ over $[0,0.09]$. }
  \label{fig:photonsphere}
\end{figure}

\begin{figure}[ht!]
  \centering
  \includegraphics[width=0.66\textwidth]{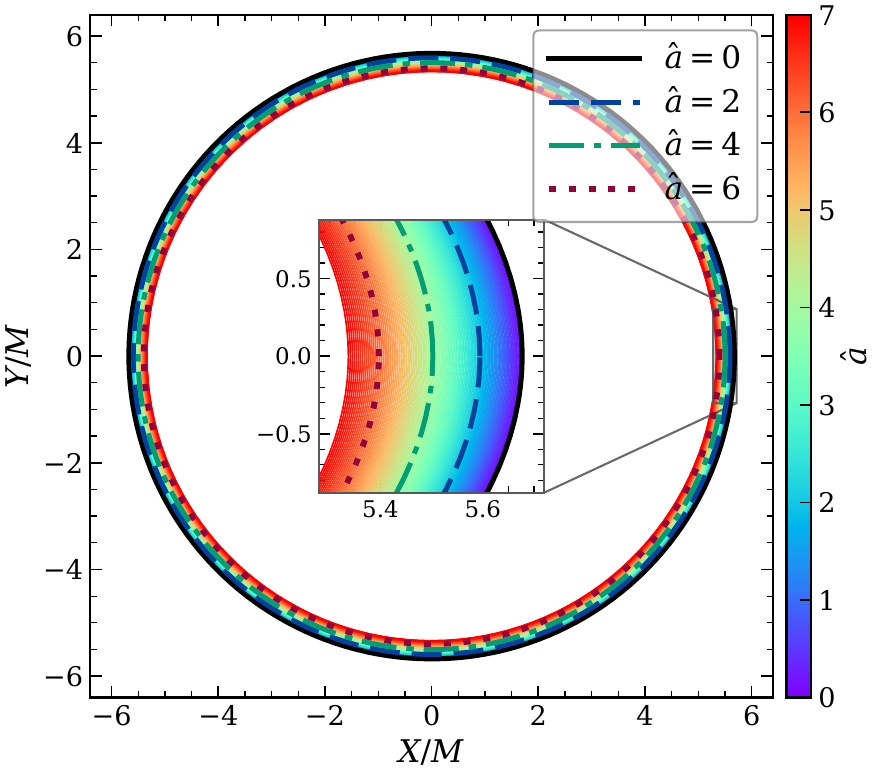}
  \caption{Shadow silhouettes of radius $b_{c}$ in the observer plane, for $\Lambda=-0.03$, $N=0.05$, $w=-2/3$, with the ribbon sweeping $\hat a$ over $[0,7]$. The inset magnifies the right-hand rim over $X/M\in[4.3,5.45]$, where successive contours are separated by a few times $10^{-2}M$.}
  \label{fig:shadow}
\end{figure}

The behaviour of the effective potential is displayed in Fig.~\ref{fig:vnull}. Raising $\hat a$ lowers the barrier and shifts its maximum inward, both of which follow from the negative sign with which the deformation enters Eq.~\eqref{eq:phsphere_explicit}. Figure~\ref{fig:photonsphere} tracks the photon sphere and the critical impact parameter together, and Table~\ref{tab:null} gives the numbers: $r_{\rm ph}$ contracts from $3.2668M$ at $\hat a=0$ to $2.8664M$ at $\hat a=7$, a reduction of about an eighth, while $b_{c}$ falls from $5.6768M$ to $5.3374M$, a reduction of about a sixteenth. The impact parameter moves less than the photon sphere because the lapse at the photon sphere rises as the orbit moves inward, and the two effects partly cancel in the ratio $r_{\rm ph}/\sqrt{f}$. The angular radius seen at $r_{o}=50M$ contracts from $33.36^{\circ}$ to $31.13^{\circ}$. Figure~\ref{fig:shadow} renders the corresponding silhouettes.

\begin{figure}[ht!]
  \centering
  \includegraphics[width=0.92\textwidth]{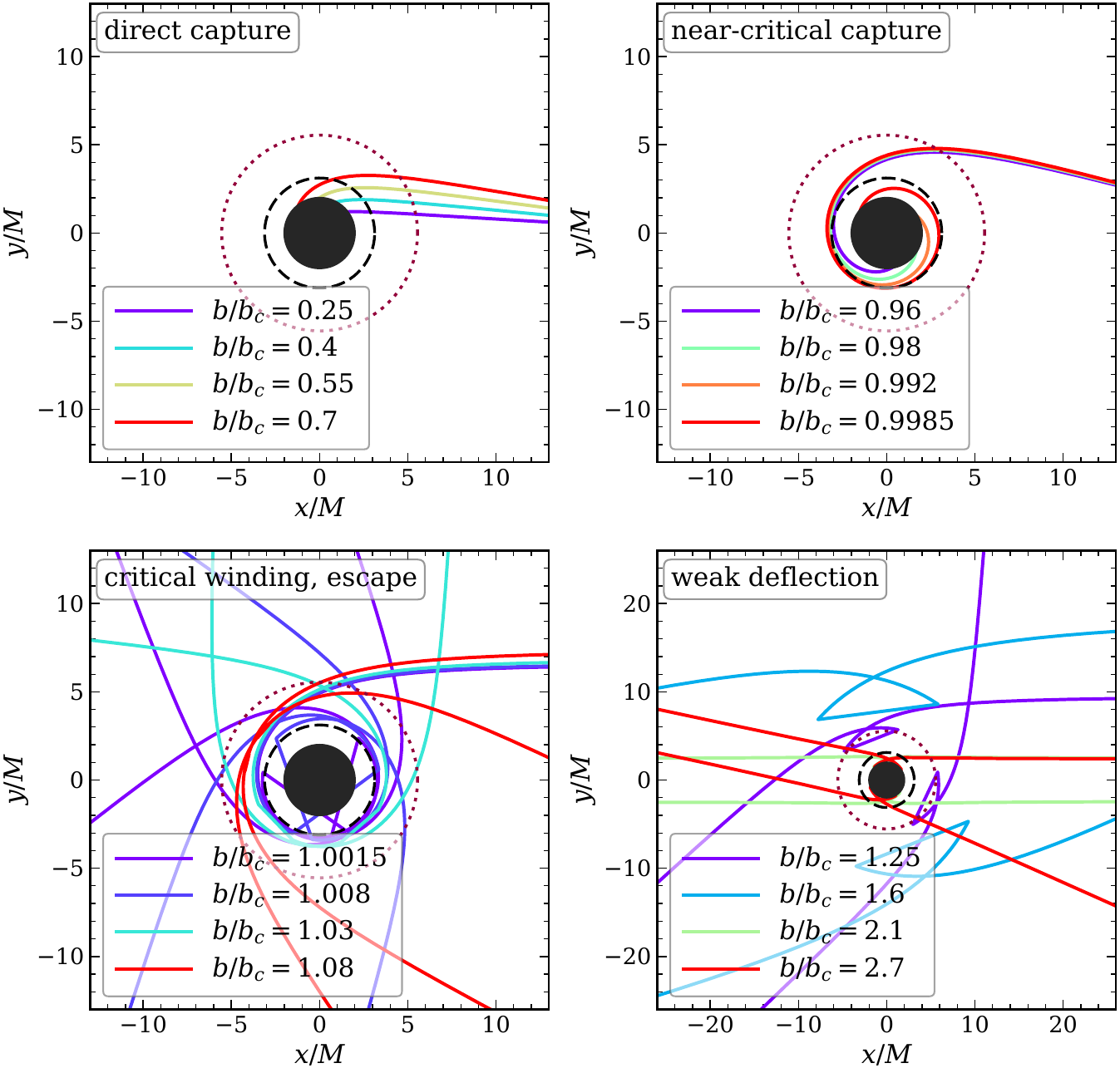}
  \caption{Null geodesics of Eq.~\eqref{eq:orbit_first} for $\hat a=3$, $\Lambda=-0.03$, $N=0.05$, $w=-2/3$, grouped by character: direct capture, near-critical capture, critical winding followed by escape, and weak deflection. The filled disc is the event horizon, the dashed circle the photon sphere and the dotted circle the critical impact parameter $b_{c}=5.55M$. Rays enter from the right; note the change of scale in the last panel.}
  \label{fig:trajectories}
\end{figure}

The bending of individual rays follows from the orbit equation. Writing $u=1/r$ and using the first integral $(du/d\phi)^{2}=b^{-2}-u^{2}f(1/u)$,
\begin{equation}
\Big(\frac{du}{d\phi}\Big)^{2}+u^{2}=\frac{1}{b^{2}}+\frac{\Lambda}{3}+2Mu^{3}-\frac{\hat aM^{3}}{2\pi}u^{5}+Nu,
\label{eq:orbit_first}
\end{equation}
for $w=-2/3$, and differentiating once,
\begin{equation}
\frac{d^{2}u}{d\phi^{2}}+u=3Mu^{2}-\frac{5\hat aM^{3}}{4\pi}u^{4}+\frac{N}{2}.
\label{eq:orbit_second}
\end{equation}
The Schwarzschild term $3Mu^{2}$ is the familiar source of perihelion advance~\cite{Jha:2026kro,Ali:2022geo}. The deformation enters at order $u^{4}$, so it is negligible in the weak-field region and takes over only near the photon sphere, while the quintessence contributes the constant $N/2$ and therefore shifts the deflection at every radius. Figure~\ref{fig:trajectories} integrates Eq.~\eqref{eq:orbit_first} directly, using the substitution $u=u_{t}-t^{2}$ about the periastron $u_{t}$ to keep the integrand finite where the square root vanishes. Rays with $b>b_{c}$ turn and escape, rays with $b<b_{c}$ spiral through the horizon, and those close to $b_{c}$ wind several times around the photon sphere before doing either. The width of the winding band is set by $\lambda_{L}$, which is why the deformation, in lowering $\lambda_{L}$, widens it.

The stability of the circular photon orbit is measured by the Lyapunov exponent of the Cardoso-Miranda-Berti-Witek-Zanchin form,
\begin{equation}
\lambda_{L}^{2}=\frac{f_{c}}{2}\Big(\frac{2f_{c}}{r_{\rm ph}^{2}}-f_{c}''\Big),\qquad f_{c}\equiv f(r_{\rm ph}),
\label{eq:lyapunov}
\end{equation}
which returns $\lambda_{L}=1/3\sqrt3 M=0.19245/M$ for Schwarzschild, as our implementation confirms. Together with the orbital frequency $\Omega_{c}=\sqrt{f_{c}}/r_{\rm ph}$ it supplies the leading eikonal estimate of the ringdown spectrum,
\begin{equation}
\omega\simeq\Omega_{c}\Big(\ell+\tfrac12\Big)-i\Big(n+\tfrac12\Big)|\lambda_{L}|.
\label{eq:eikonal}
\end{equation}
Table~\ref{tab:null} shows that $\Omega_{c}$ rises with $\hat a$ while $\lambda_{L}$ falls, so the deformation raises the oscillation frequency and lengthens the damping time at the same time. Table~\ref{tab:qnm} converts this into frequencies for $\ell=2,3,4$ and $n=0,1$. Two cautions apply to those numbers. Equation~\eqref{eq:eikonal} is a leading-order result in $1/\ell$ and is not universal; it can fail for gravitational and non-minimally coupled perturbations in modified gravity, where the peak of the potential decouples from the null circular orbit. And the AdS boundary imposes its own conditions on the wave equation that the eikonal formula does not see. The entries in Table~\ref{tab:qnm} should therefore be read as photon-sphere data expressed in frequency units, not as a solution of the perturbation problem.

\begin{figure}[ht!]
  \centering
  \includegraphics[width=0.78\textwidth]{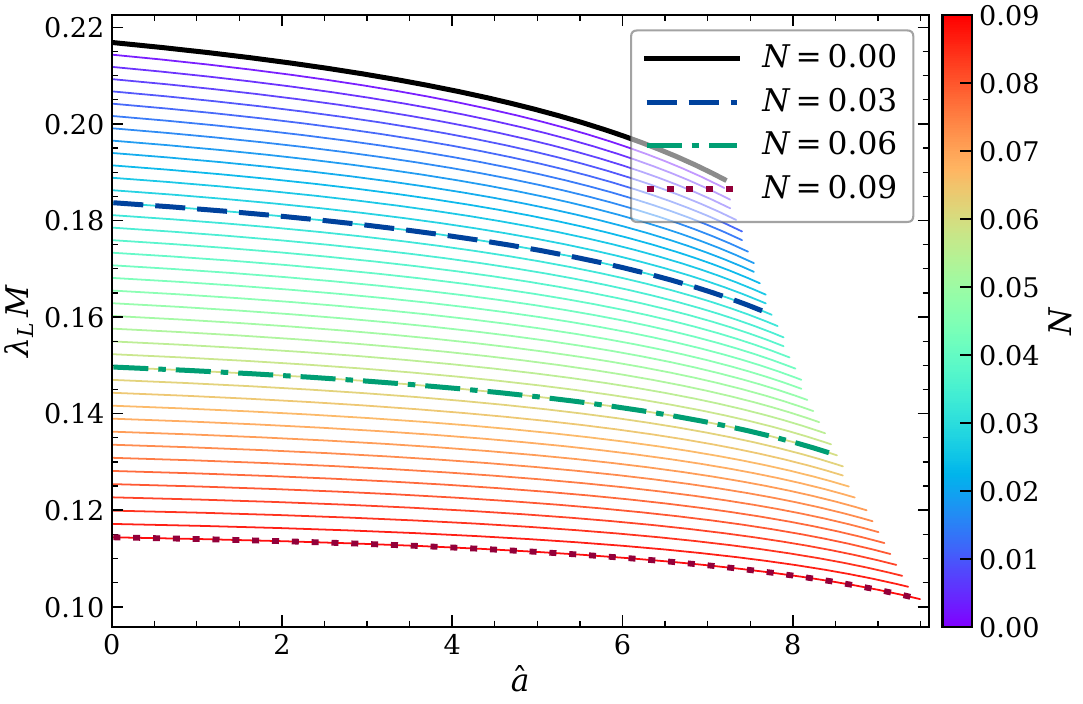}
  \caption{Lyapunov exponent of Eq.~\eqref{eq:lyapunov} against $\hat a$, for $\Lambda=-0.03$, $w=-2/3$, with the ribbon sweeping $N$ over $[0,0.09]$. Each curve terminates where its horizon ceases to exist, at $\hat a=7.20$ for $N=0$ and $\hat a=9.49$ for $N=0.09$.}
  \label{fig:lyapunov}
\end{figure}

\begin{longtable}{L{0.0713\textwidth} L{0.1110\textwidth} L{0.1110\textwidth} L{0.1189\textwidth} L{0.1189\textwidth} L{0.1268\textwidth}}
  \hline\hline
\cellcolor{brown!30}$\hat a$ & \cellcolor{brown!30}$r_{\rm ph}/M$ & \cellcolor{brown!30}$b_{c}/M$ & \cellcolor{brown!30}$\Omega_{c}M$ & \cellcolor{brown!30}$\lambda_{L}M$ & \cellcolor{brown!30}$\alpha_{\rm sh}$ (deg) \\
  \hline
  \endfirsthead
  \endhead
  \hline
  \endfoot
  \hline\hline
  \caption{Null-geodesic observables for $M=1$, $\Lambda=-0.03$, $N=0.05$, $w=-2/3$. The angular radius follows from Eq.~\eqref{eq:shadow_angle} with the static observer at $r_{o}=50M$; in an AdS background this angle depends on $r_{o}$ and the value must be quoted with it.}
  \label{tab:null}\\
  \endlastfoot
0.0 & 3.26680 & 5.67680 & 0.17616 & 0.16113 & 33.3614 \\
1.0 & 3.22102 & 5.63670 & 0.17741 & 0.16017 & 33.0953 \\
2.0 & 3.17257 & 5.59450 & 0.17875 & 0.15903 & 32.8162 \\
3.0 & 3.12096 & 5.54989 & 0.18018 & 0.15764 & 32.5220 \\
4.0 & 3.06559 & 5.50247 & 0.18174 & 0.15594 & 32.2104 \\
5.0 & 3.00562 & 5.45176 & 0.18343 & 0.15381 & 31.8784 \\
6.0 & 2.93984 & 5.39707 & 0.18529 & 0.15111 & 31.5216 \\
7.0 & 2.86643 & 5.33744 & 0.18736 & 0.14754 & 31.1341 \\
\end{longtable}

\begin{longtable}{L{0.0392\textwidth} L{0.0392\textwidth} L{0.1449\textwidth} L{0.1449\textwidth} L{0.1449\textwidth} L{0.1449\textwidth}}
  \hline\hline
  \cellcolor{brown!30}$\ell$ & \cellcolor{brown!30}$n$ & \cellcolor{brown!30}$\hat a=0$ & \cellcolor{brown!30}$\hat a=2$ & \cellcolor{brown!30}$\hat a=4$ & \cellcolor{brown!30}$\hat a=6$ \\
  \hline
  \endfirsthead
  \endhead
  \hline
  \endfoot
  \hline\hline
  \caption{Eikonal ringdown frequencies $\omega M$ from Eq.~\eqref{eq:eikonal}, for $\Lambda=-0.03$, $N=0.05$, $w=-2/3$. These are photon-sphere data expressed in frequency units at leading order in $1/\ell$; they are not a solution of the perturbation problem with anti-de Sitter boundary conditions, and the correspondence they rest on is not universal.}
  \label{tab:qnm}\\
  \endlastfoot
2 & 0 & 0.44039$-$0.08056$i$ & 0.44687$-$0.07951$i$ & 0.45434$-$0.07797$i$ & 0.46321$-$0.07555$i$ \\
  2 & 1 & 0.44039$-$0.24169$i$ & 0.44687$-$0.23854$i$ & 0.45434$-$0.23391$i$ & 0.46321$-$0.22666$i$ \\
  3 & 0 & 0.61654$-$0.08056$i$ & 0.62561$-$0.07951$i$ & 0.63608$-$0.07797$i$ & 0.64850$-$0.07555$i$ \\
  3 & 1 & 0.61654$-$0.24169$i$ & 0.62561$-$0.23854$i$ & 0.63608$-$0.23391$i$ & 0.64850$-$0.22666$i$ \\
  4 & 0 & 0.79270$-$0.08056$i$ & 0.80436$-$0.07951$i$ & 0.81781$-$0.07797$i$ & 0.83379$-$0.07555$i$ \\
  4 & 1 & 0.79270$-$0.24169$i$ & 0.80436$-$0.23854$i$ & 0.81781$-$0.23391$i$ & 0.83379$-$0.22666$i$ \\
\end{longtable}

\section{Timelike geodesics, circular orbits and epicyclic frequencies}\label{isec4}

Massive test particles probe the geometry at radii well outside the photon sphere, which is where accretion physics lives and where the deformation competes against the quintessence tail rather than dominating it. Normalising the four-velocity gives $\dot r^{2}=E^{2}-V_{\rm eff}$ with
\begin{equation}
V_{\rm eff}(r)=f(r)\Big(1+\frac{L^{2}}{r^{2}}\Big),
\label{eq:vtimelike}
\end{equation}
and circular orbits at radius $r_{c}$ require $V_{\rm eff}=E^{2}$ together with $V'_{\rm eff}=0$. Solving the pair gives
\begin{equation}
L^{2}=\frac{r^{3}f'(r)}{2f(r)-rf'(r)},\qquad
E^{2}=\frac{2f(r)^{2}}{2f(r)-rf'(r)},
\label{eq:EL_circ}
\end{equation}
both of which require $2f-rf'>0$. The locus where that denominator vanishes is precisely the photon sphere of Eq.~\eqref{eq:phsphere}, so circular timelike orbits are confined to the region between the inner photon sphere and the outer one that the quintessence tail creates. Outside $r\simeq2/N$ the expressions in Eq.~\eqref{eq:EL_circ} change sign and no circular orbit exists, which is the reason the marginally bound orbit is not a useful reference radius here: with $\Lambda<0$ the potential rises without bound at large $r$ and every timelike geodesic is bound in any case.

The innermost stable circular orbit is the marginally stable member of the family, fixed by $d(E^{2})/dr=0$, equivalently by $V''_{\rm eff}=0$ imposed alongside the two circular-orbit conditions. It should not be read off $f''=0$. Equivalently, imposing $U''_{\rm eff}\ge0$ on the circular family and taking the marginal case gives the compact condition
\begin{equation}
f(r)f''(r)-2\big(f'(r)\big)^{2}+\frac{3f(r)f'(r)}{r}=0,
\label{eq:mso}
\end{equation}
which divides the numerator of $d(E^{2})/dr$ exactly and therefore has the innermost stable circular orbit among its roots. We solve Eq.~\eqref{eq:mso} numerically and confirm that its residual vanishes at the radius returned by $d(E^{2})/dr=0$ to better than $10^{-11}$ for every parameter set used below.

Our implementation returns $r_{\rm ISCO}=6M$, $E=2\sqrt2/3$ and $L=2\sqrt3 M$ in the Schwarzschild limit. Decomposing the baseline configuration is instructive, because the three ingredients pull in different directions. Taking them one at a time with $M=1$: Schwarzschild gives $6M$; adding $\Lambda=-0.03$ alone contracts the orbit to $4.291M$; adding $N=0.05$, $w=-2/3$ alone expands it to $17.746M$; and the two together give $4.607M$, so the AdS term wins. Switching on $\hat a=4$ then contracts it further to $4.398M$. The deformation and the cosmological constant therefore act in the same direction on the ISCO, and against the quintessence.

\begin{figure}[ht!]
  \centering
  \includegraphics[width=0.78\textwidth]{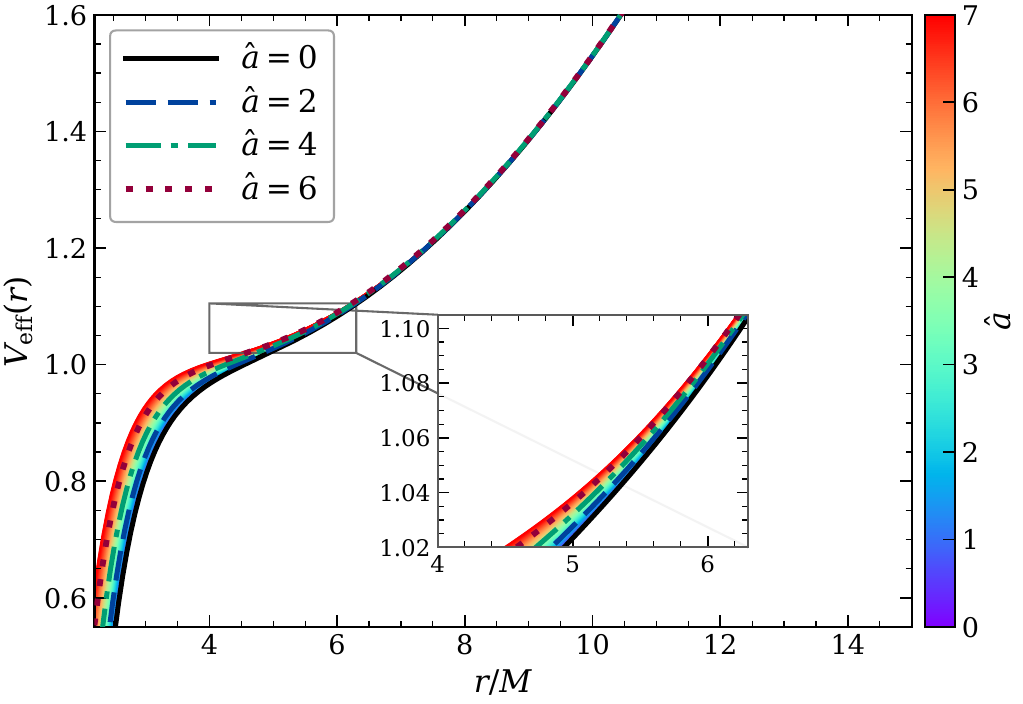}
  \caption{Timelike effective potential of Eq.~\eqref{eq:vtimelike} at fixed $L=4.2M$, for $\Lambda=-0.03$, $N=0.05$, $w=-2/3$, with the ribbon sweeping $\hat a$ over $[0,7]$. The inset covers $r/M\in[4.0,6.3]$ near the potential minimum, where the curves lie within about $10^{-2}$ of one another.}
  \label{fig:vtimelike}
\end{figure}

\begin{figure}[ht!]
  \centering
  \includegraphics[width=0.78\textwidth]{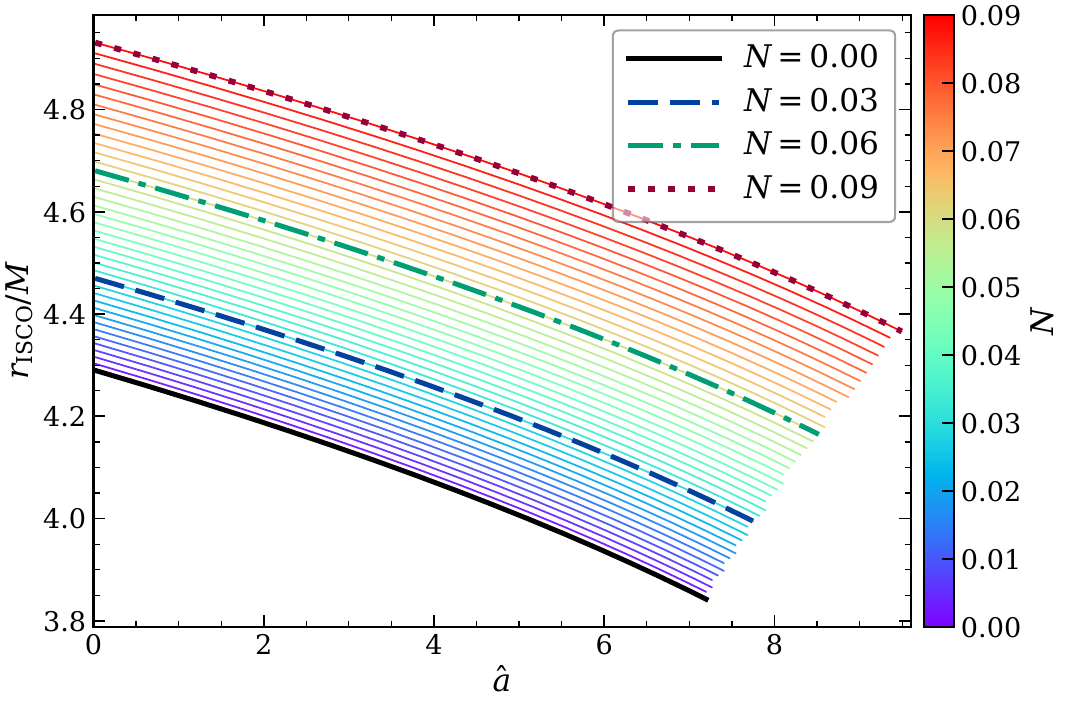}
  \caption{Innermost stable circular orbit against $\hat a$, for $\Lambda=-0.03$, $w=-2/3$, with the ribbon sweeping $N$ over $[0,0.09]$. Each curve terminates where its horizon ceases to exist.}
  \label{fig:isco}
\end{figure}

Table~\ref{tab:timelike} collects the orbital data. The ISCO contracts steadily with $\hat a$, from $4.6072M$ to $4.2021M$ over the admissible range, and the specific energy at that orbit falls from $1.1333$ to $1.0814$ while the specific angular momentum falls from $5.3419M$ to $4.8152M$. The energies exceed unity because the AdS potential lifts the whole effective potential; this is a property of the background rather than of the deformation, and it is present already at $\hat a=0$. The orbital frequency at the ISCO rises with $\hat a$, from $0.1217/M$ to $0.1275/M$, consistent with the orbit moving inward. The outer photon orbit listed in the second column sits at $36.73M$ and is almost completely insensitive to the deformation, which is expected: at that radius the $r^{-3}$ term is suppressed by roughly five orders of magnitude relative to the quintessence tail. Figure~\ref{fig:vtimelike} shows the effective potential and Fig.~\ref{fig:isco} the ISCO trend.

\begin{figure}[ht!]
  \centering
  \includegraphics[width=0.78\textwidth]{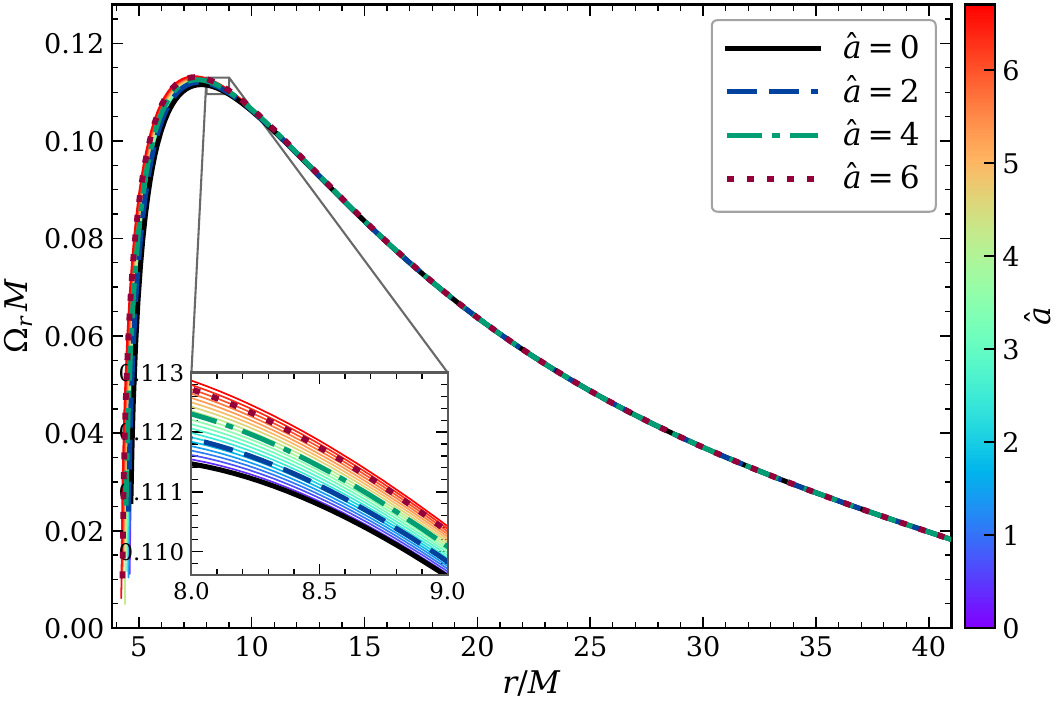}
  \caption{Radial epicyclic frequency of Eq.~\eqref{eq:epicyclic} for $\Lambda=-0.03$, $N=0.05$, $w=-2/3$, with the ribbon sweeping $\hat a$ over $[0,7]$. The inset covers $r/M\in[8,9]$ near the maximum, where the curves are separated by less than $4\times10^{-4}$.}
  \label{fig:epicyclic}
\end{figure}

Small perturbations of a circular orbit oscillate at the radial epicyclic frequency
\begin{equation}
\Omega_{r}^{2}=\frac{f(r)}{2E^{2}}\,\frac{d^{2}V_{\rm eff}}{dr^{2}},
\label{eq:epicyclic}
\end{equation}
while the vertical epicyclic frequency degenerates with the orbital frequency $\Omega_{\phi}=\sqrt{f'(r)/2r}$ because the metric is spherically symmetric; rotation is what lifts that degeneracy, and there is none here. The ISCO is the radius at which $\Omega_{r}$ vanishes, and our implementation reproduces this to four decimal places in the Schwarzschild case. Figure~\ref{fig:epicyclic} shows $\Omega_{r}$ across the orbital region. The frequency rises from zero at the ISCO, peaks, and falls back toward zero as the outer photon orbit is approached, so the region supporting stable circular orbits is bounded on both sides. Raising $\hat a$ shifts the whole profile inward and slightly upward, following the ISCO. For a quasiperiodic-oscillation model this matters through the ratio $\Omega_{\phi}/\Omega_{r}$ rather than through either frequency alone, and since the deformation moves the ISCO by only a few per cent over its entire admissible range, twin-peak frequencies constrain $\hat a$ far less tightly than the shadow does.

\begin{longtable}{L{0.0713\textwidth} L{0.1268\textwidth} L{0.1189\textwidth} L{0.1110\textwidth} L{0.1110\textwidth} L{0.1189\textwidth}}
  \hline\hline
\cellcolor{brown!30}$\hat a$ & \cellcolor{brown!30}$r_{\rm ph}^{\rm out}/M$ & \cellcolor{brown!30}$r_{\rm ISCO}/M$ & \cellcolor{brown!30}$E_{\rm ISCO}$ & \cellcolor{brown!30}$L_{\rm ISCO}/M$ & \cellcolor{brown!30}$\Omega_{\phi}M$ \\
  \hline
  \endfirsthead
  \endhead
  \hline
  \endfoot
  \hline\hline
  \caption{Timelike-orbit observables for $M=1$, $\Lambda=-0.03$, $N=0.05$, $w=-2/3$. The second column is the outer root of Eq.~\eqref{eq:phsphere_explicit} produced by the quintessence tail; circular timelike orbits exist only between it and the inner photon sphere of Table~\ref{tab:null}. The last three columns are evaluated at the ISCO.}
  \label{tab:timelike}\\
  \endlastfoot
0.0 & 36.73320 & 4.60719 & 1.13325 & 5.34192 & 0.12165 \\
1.0 & 36.73355 & 4.55882 & 1.12665 & 5.27377 & 0.12227 \\
2.0 & 36.73391 & 4.50803 & 1.11983 & 5.20369 & 0.12294 \\
3.0 & 36.73426 & 4.45448 & 1.11278 & 5.13149 & 0.12367 \\
4.0 & 36.73461 & 4.39774 & 1.10544 & 5.05690 & 0.12447 \\
5.0 & 36.73496 & 4.33728 & 1.09780 & 4.97960 & 0.12536 \\
6.0 & 36.73531 & 4.27237 & 1.08979 & 4.89919 & 0.12635 \\
7.0 & 36.73567 & 4.20207 & 1.08136 & 4.81515 & 0.12749 \\
\end{longtable}

\section{Thermodynamics and the corrected entropy}\label{isec5}

The thermodynamics of this solution is where the mass-dependence of the deformation amplitude stops being a technical nuisance and starts producing physics. We derive the temperature, show that the area law and the first law cannot both hold, correct the entropy accordingly, and check the outcome numerically.

\begin{figure}[ht!]
  \centering
  \includegraphics[width=0.8\textwidth]{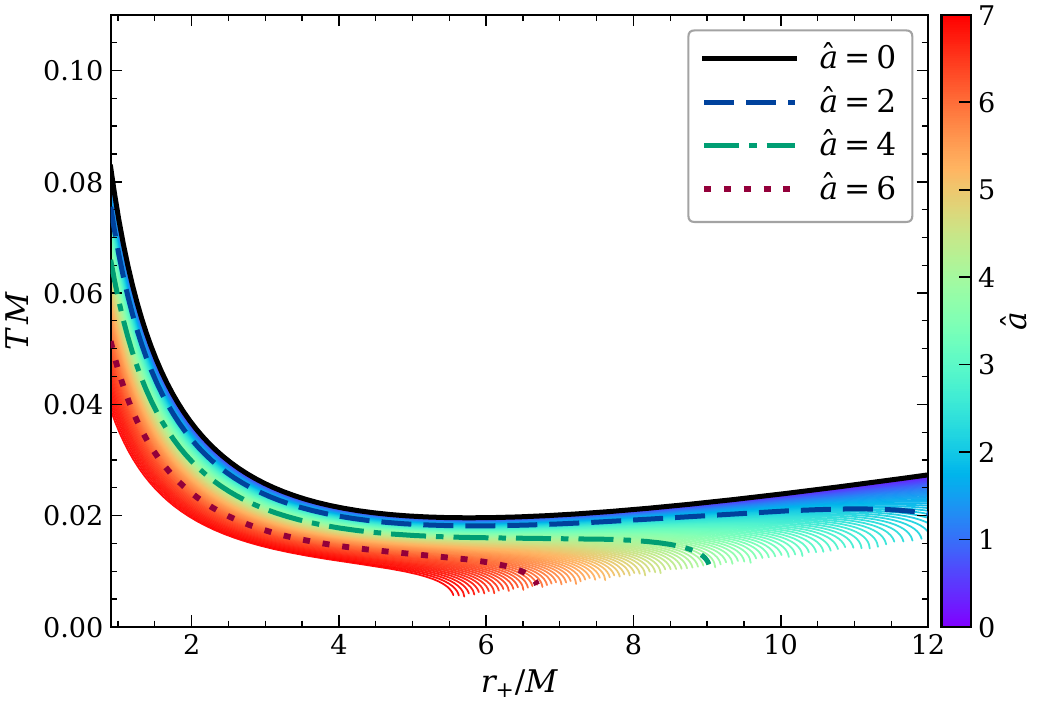}
  \caption{Hawking temperature of Eq.~\eqref{eq:temperature} for $\Lambda=-0.03$, $N=0.05$, $w=-2/3$, with the ribbon sweeping $\hat a$ over $[0,7]$. Curves terminate where the admissibility bound of Eq.~\eqref{eq:amax} is reached. The inset covers $r_{+}/M\in[2.2,6]$ around the temperature minimum.}
  \label{fig:temperature}
\end{figure}

Surface gravity fixes the temperature~\cite{Bardeen:1973gs,Hawking:1974rv,Hawking:1975vcx}, and the area law and the generalized second law give the entropy its usual meaning~\cite{Bekenstein:1972tm,Bekenstein:1973ur,Bekenstein:1974ax,Hawking:1976de}. Since $g_{tt}g_{rr}=-1$ the standard expression applies,
\begin{equation}
T=\frac{f'(r_{+})}{4\pi}
=\frac{1}{4\pi r_{+}}-\frac{\Lambda r_{+}}{4\pi}+\frac{3wN}{4\pi r_{+}^{3w+2}}-\frac{\hat a M^{3}}{4\pi^{2}r_{+}^{4}},
\label{eq:temperature}
\end{equation}
where the second equality follows from eliminating $2M/r_{+}$ with the horizon condition and where $M$ is understood as the branch of Eq.~\eqref{eq:mass_branch}. The four terms are transparent. The first is the Schwarzschild contribution; with $\hat a=N=\Lambda=0$ and $r_{+}=2M$ it returns $T=1/8\pi M$. The second is the AdS contribution, positive for $\Lambda<0$, which is what makes large AdS black holes hot. The third carries the quintessence, and since $w<0$ for quintessence proper it lowers the temperature. The fourth is the deformation, and it lowers the temperature as well. Figure~\ref{fig:temperature} shows the result: the deformation depresses the whole $T(r_{+})$ profile and steepens the small-$r_{+}$ branch, so a deformed black hole of given horizon radius is colder than its undeformed counterpart. At $r_{+}=3M$ the temperature falls from $0.02573/M$ at $\hat a=0$ to $0.01438/M$ at $\hat a=7$, a reduction of nearly a half.

Now consider the first law. Writing the horizon condition as $h(M,r_{+})=0$ with $h=f(r_{+})$ and differentiating implicitly,
\begin{equation}
\frac{dM}{dr_{+}}=-\frac{\partial h/\partial r_{+}}{\partial h/\partial M}
=\frac{f'(r_{+})}{\dfrac{2}{r_{+}}-\dfrac{3\hat a M^{2}}{2\pi r_{+}^{3}}}.
\label{eq:dMdr}
\end{equation}
The denominator is the point. For $\hat a=0$ it is $2/r_{+}$, and $dM/dr_{+}=r_{+}f'/2$, which combined with $dS=2\pi r_{+}dr_{+}$ delivers $dM=T\,dS$ with $S=\pi r_{+}^{2}$ in the usual way. For $\hat a>0$ the extra term spoils that cancellation. Since the temperature is fixed by the surface gravity and is not negotiable, the first law can be preserved only by correcting the entropy. Demanding $dM=T\,dS$ and using Eqs.~\eqref{eq:temperature} and~\eqref{eq:dMdr},
\begin{equation}
\frac{dS}{dr_{+}}=\frac{1}{T}\frac{dM}{dr_{+}}=\frac{2\pi r_{+}^{3}}{r_{+}^{2}-\dfrac{3\hat a M(r_{+})^{2}}{4\pi}},
\label{eq:dSdr}
\end{equation}
so that
\begin{equation}
S(r_{+})=\int^{r_{+}}\frac{2\pi r^{3}\,dr}{r^{2}-\dfrac{3\hat a M(r)^{2}}{4\pi}},
\label{eq:entropy_exact}
\end{equation}
which reduces to $\pi r_{+}^{2}$ when $\hat a=0$ and which is larger than the area law whenever $\hat a>0$, since the denominator is then smaller than $r_{+}^{2}$. Expanding to first order in $\hat a$ with $M\to r_{+}G(r_{+})/2$ and integrating term by term gives the closed form
\begin{equation}
S=\pi r_{+}^{2}+\frac{3\hat a}{8}\left[\frac{r_{+}^{2}}{2}-\frac{\Lambda r_{+}^{4}}{6}+\frac{\Lambda^{2}r_{+}^{6}}{54}-\frac{N^{2}r_{+}^{-6w}}{6w}-\frac{2Nr_{+}^{1-3w}}{1-3w}+\frac{2\Lambda Nr_{+}^{3-3w}}{9(1-w)}\right]+\mathcal{O}(\hat a^{2}),
\label{eq:entropy_pert}
\end{equation}
the integration constant being fixed by requiring $S\to0$ as $r_{+}\to0$ in the $\Lambda=N=0$ case. In that Schwarzschild sector Eq.~\eqref{eq:entropy_pert} collapses to
\begin{equation}
S=\pi r_{+}^{2}\Big(1+\frac{3\hat a}{16\pi}\Big),
\label{eq:entropy_schw}
\end{equation}
so the deformation rescales the area law by a constant factor rather than adding a logarithm. That is a mildly surprising outcome, since minimal-length corrections usually announce themselves through a $\log A$ term; here the correction is a pure rescaling because the deformation and the mass scale together in exactly the way that keeps the ratio $S/A$ radius-independent.

\begin{figure}[ht!]
  \centering
  \includegraphics[width=0.8\textwidth]{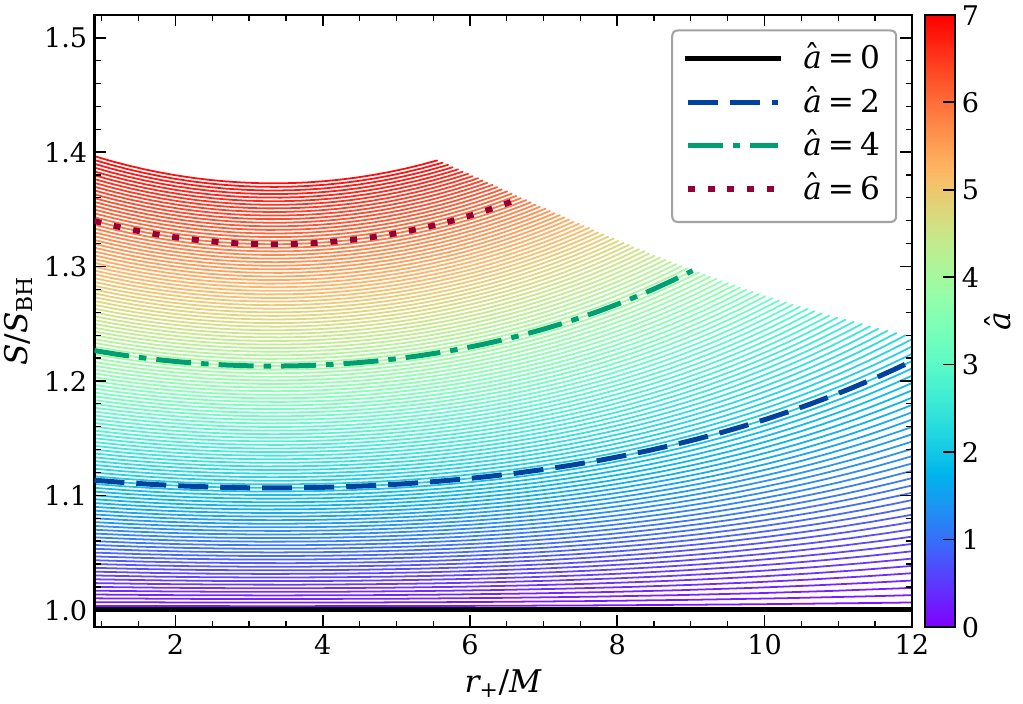}
  \caption{Ratio of the corrected entropy of Eq.~\eqref{eq:entropy_exact} to the Bekenstein-Hawking value $S_{\rm BH}=\pi r_{+}^{2}$, for $\Lambda=-0.03$, $N=0.05$, $w=-2/3$, with the ribbon sweeping $\hat a$ over $[0,7]$. The horizontal line at unity marks the area law. The inset covers $r_{+}/M\in[7.5,12]$, where the low-$\hat a$ curves crowd within one per cent of unity.}
  \label{fig:entropy}
\end{figure}

Table~\ref{tab:entropy} compares the exact quadrature of Eq.~\eqref{eq:entropy_exact} against the first-order form of Eq.~\eqref{eq:entropy_pert}. The two agree to one part in $10^{3}$ at $\hat a=0.5$ and to about one part in ten at $\hat a=4$, which sets the range over which the closed form can be used in place of the integral. Table~\ref{tab:firstlaw} is the consistency check that matters: across six values of $\hat a$ and three horizon radii, $dM/dS$ and $T$ agree to between $10^{-12}$ and $10^{-10}$, confirming that Eq.~\eqref{eq:entropy_exact} is the entropy the first law selects. Figure~\ref{fig:entropy} plots the ratio $S/S_{\rm BH}$, which grows monotonically with $\hat a$ and reaches $1.34$ at $\hat a=4$, $r_{+}=3M$, and $2.33$ at $\hat a=7$.

There is a second, equivalent way to restore consistency, and setting the two side by side clarifies what the deformation actually does. Instead of correcting the entropy one may keep $S=\pi r_{+}^{2}$ and rescale the mass, following the corrected first law devised for regular black holes~\cite{Ma:2014qma} and applied since to noncommutative geometries~\cite{Kubiznak:2017nce,Ma:2017pvc}, replacing the first law by
\begin{equation}
d\mathcal{M}=W\,dM=T\,dS+V\,dP+\Phi_{\hat a}\,d\hat a+\Phi_{N}\,dN,
\label{eq:modified_first_law}
\end{equation}
where the correction factor is fixed by the response of the source to a change of mass,
\begin{equation}
W=1+4\pi\int_{r_{+}}^{\infty}r^{2}\,\frac{\partial T^{0}{}_{0}}{\partial M}\,dr.
\label{eq:W_def}
\end{equation}
The integral is evaluated from the field equation $\partial_{r}(rf)-1+r^{2}\Lambda=8\pi r^{2}T^{0}{}_{0}$. Differentiating with respect to $M$ turns the integrand into a total derivative,
\begin{equation}
\int_{r_{+}}^{\infty}8\pi r^{2}\frac{\partial T^{0}{}_{0}}{\partial M}\,dr
=\Big[r\frac{\partial f}{\partial M}\Big]_{r_{+}}^{\infty},
\end{equation}
and with $\partial f/\partial M=-2/r+3\hat aM^{2}/2\pi r^{3}$ the boundary terms give
\begin{equation}
W=1-\frac{3\hat aM^{2}}{4\pi r_{+}^{2}}.
\label{eq:W}
\end{equation}
The two prescriptions are the same statement written twice, because Eq.~\eqref{eq:dSdr} is exactly
\begin{equation}
\frac{dS}{dr_{+}}=\frac{2\pi r_{+}}{W},
\label{eq:dS_W}
\end{equation}
which we verified symbolically and numerically. Correcting the entropy by $1/W$ and rescaling the mass by $W$ move the same factor between the two sides of $dM=T\,dS$.

\begin{figure}[ht!]
  \centering
  \includegraphics[width=0.8\textwidth]{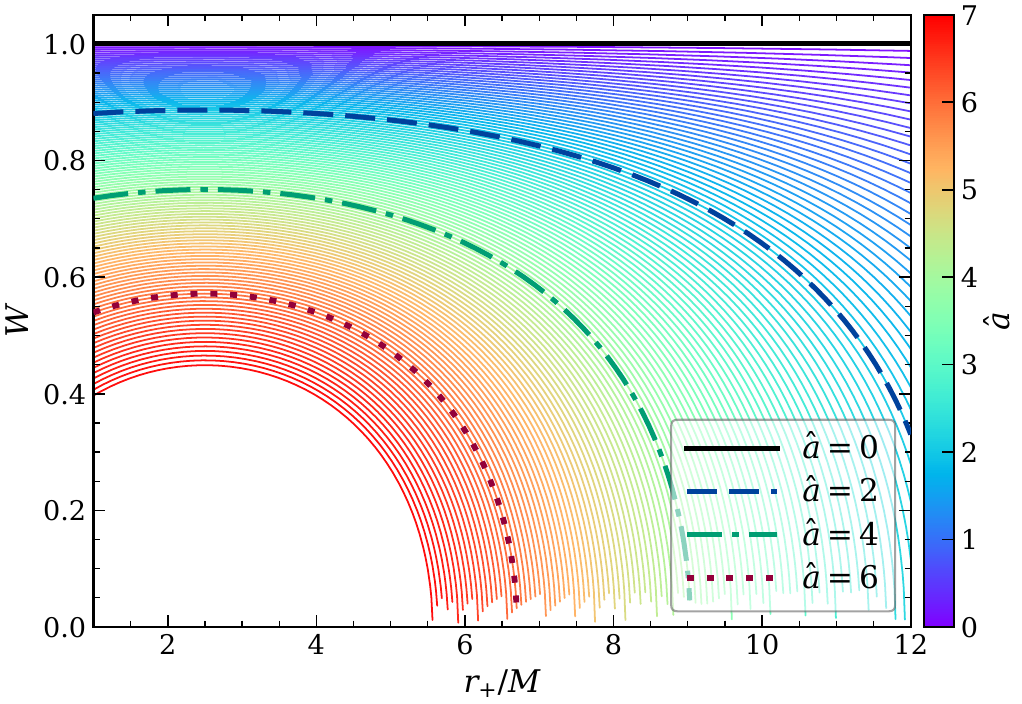}
  \caption{Correction factor of Eq.~\eqref{eq:W} for $\Lambda=-0.03$, $N=0.05$, $w=-2/3$, with the ribbon sweeping $\hat a$ over $[0,7]$. The horizontal line at unity is the undeformed value. Curves terminate where the admissibility bound of Eq.~\eqref{eq:amax} is reached.}
  \label{fig:Wfactor}
\end{figure}

Working in the rescaled frame, the conjugates follow from differentiating the horizon condition at fixed $S$, and every one of them comes out clean:
\begin{align}
T&=W\Big(\frac{\partial M}{\partial S}\Big)_{P,\hat a,N}=\frac{f'(r_{+})}{4\pi}=T_{H},
&V&=W\Big(\frac{\partial M}{\partial P}\Big)_{S,\hat a,N}=\frac{4\pi r_{+}^{3}}{3},
\label{eq:TV}\\
\Phi_{\hat a}&=W\Big(\frac{\partial M}{\partial\hat a}\Big)_{S,P,N}=\frac{M^{3}}{4\pi r_{+}^{2}},
&\Phi_{N}&=W\Big(\frac{\partial M}{\partial N}\Big)_{S,P,\hat a}=-\frac{r_{+}^{-3w}}{2}.
\label{eq:conjugates}
\end{align}
The temperature reproduces the surface-gravity value and the volume reproduces the geometric one, neither of which is guaranteed in advance; that they do is the check that Eq.~\eqref{eq:W} is the right factor. For $w=-2/3$ the quintessence potential is $\Phi_{N}=-r_{+}^{2}/2$. Applying the Euler scaling theorem~\cite{Li:2018gmn,Sakalli:2025lqg}, with $\hat a$ dimensionless and $[N]=L^{3w+1}$, gives the Smarr relation
\begin{equation}
\mathcal{M}=WM=2\big(TS-PV\big)+(3w+1)\,N\Phi_{N},
\label{eq:smarr}
\end{equation}
which we confirmed symbolically. No $\hat a\Phi_{\hat a}$ term appears, precisely because $\hat a$ carries no length dimension; the deformation is felt in the Smarr relation only through $W$ on the left-hand side. Figure~\ref{fig:Wfactor} shows $W$ across the parameter range. It falls from unity as the deformation grows and as the horizon shrinks, reaching $0.65$ at $\hat a=5$, $r_{+}=4M$, so the rescaling is not a small correction at the upper end of the admissible window.

A third reading is available and should be set aside explicitly. If one treats the combination $\hat aM^{3}/2\pi$ as an independent hair held fixed while the mass varies, then $\partial h/\partial M=-2/r_{+}$, the denominator in Eq.~\eqref{eq:dMdr} reverts to its undeformed value, $W=1$, and the area law satisfies the first law with no correction at all. That reading is internally consistent but attributes to the black hole a free parameter the construction of Ref.~\cite{Kumara:2026uwi} does not supply: $\hat a$ is a property of the deformed spacetime, not of the solution, so it is the coefficient $\hat aM^{3}/2\pi$ and not $\hat a$ that changes as the hole evaporates. We therefore keep $W\neq1$ throughout, using Eq.~\eqref{eq:entropy_exact} when the entropy itself is wanted and Eq.~\eqref{eq:modified_first_law} when the conjugates are. Every geodesic result of Secs.~\ref{isec3} and~\ref{isec4} is untouched by the choice.

\begin{figure}[ht!]
  \centering
  \includegraphics[width=0.8\textwidth]{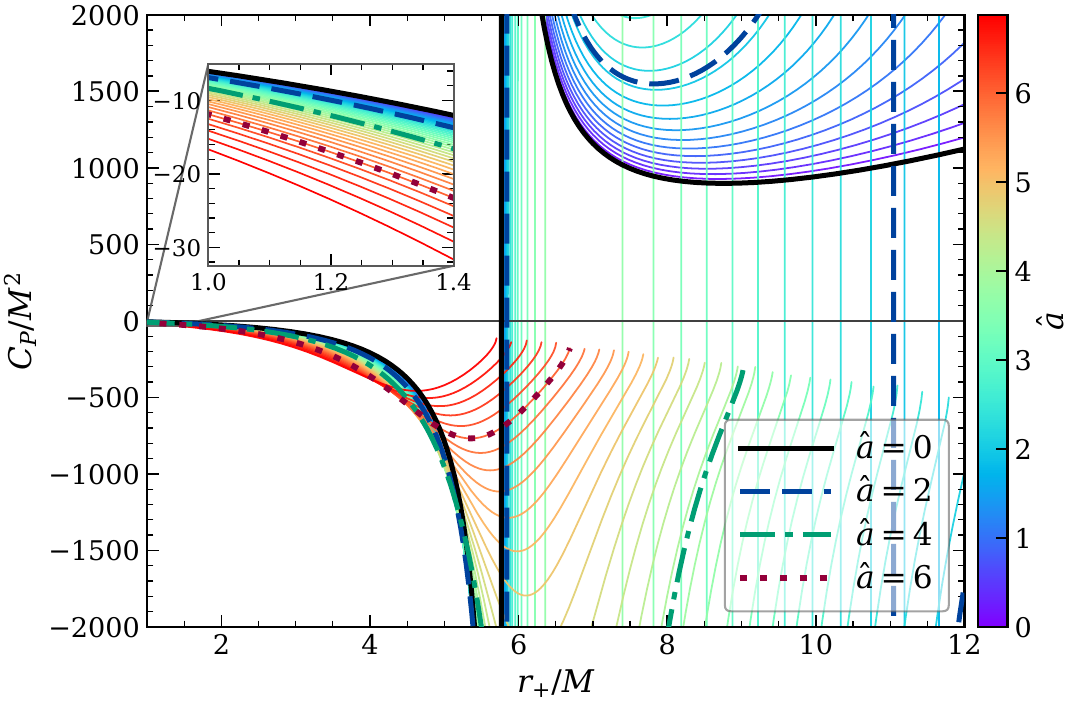}
  \caption{Heat capacity at fixed pressure for $\Lambda=-0.03$, $N=0.05$, $w=-2/3$, with the ribbon sweeping $\hat a$ over $[0,7]$. The inset covers $r_{+}/M\in[1,1.4]$, resolving the small negative branch that the full vertical scale compresses.}
  \label{fig:heatcapacity}
\end{figure}

The heat capacity at fixed pressure follows from $C_{P}=T(dS/dr_{+})(dT/dr_{+})^{-1}$ with Eq.~\eqref{eq:dSdr} supplying the numerator. Figure~\ref{fig:heatcapacity} shows it. The divergence separating the small unstable branch from the large stable branch survives the deformation and moves inward as $\hat a$ grows, tracking the shift of the temperature maximum. Table~\ref{tab:thermo} tabulates the thermodynamic quantities at $r_{+}=3M$; the heat capacity there is negative for every $\hat a$, placing that radius on the unstable branch throughout, and its magnitude grows from $75.1M^{2}$ to $183.2M^{2}$ as the deformation is switched on.

\begin{longtable}{L{0.0713\textwidth} L{0.1110\textwidth} L{0.1110\textwidth} L{0.1189\textwidth} L{0.1189\textwidth} L{0.1268\textwidth}}
  \hline\hline
\cellcolor{brown!30}$\hat a$ & \cellcolor{brown!30}$M$ & \cellcolor{brown!30}$TM$ & \cellcolor{brown!30}$S/M^{2}$ & \cellcolor{brown!30}$S/S_{\rm BH}$ & \cellcolor{brown!30}$C_{P}/M^{2}$ \\
  \hline
  \endfirsthead
  \endhead
  \hline
  \endfoot
  \hline\hline
  \caption{Thermodynamic quantities at fixed horizon radius $r_{+}=3M$, for $\Lambda=-0.03$, $N=0.05$, $w=-2/3$. The entropy is the exact quadrature of Eq.~\eqref{eq:entropy_exact}; $S_{\rm BH}=\pi r_{+}^{2}=28.2743$. The heat capacity is negative throughout, so this radius lies on the unstable branch for every admissible $\hat a$.}
  \label{tab:thermo}\\
  \endlastfoot
0.0 & 1.41000 & 0.02573 & 28.2743 & 1.00000 & $-75.140$ \\
1.0 & 1.43619 & 0.02480 & 29.9304 & 1.05857 & $-80.077$ \\
2.0 & 1.46568 & 0.02376 & 31.9600 & 1.13035 & $-86.166$ \\
3.0 & 1.49942 & 0.02257 & 34.5306 & 1.22127 & $-93.922$ \\
4.0 & 1.53889 & 0.02117 & 37.9386 & 1.34180 & $-104.242$ \\
5.0 & 1.58656 & 0.01949 & 42.7748 & 1.51285 & $-118.855$ \\
6.0 & 1.64703 & 0.01735 & 50.4583 & 1.78460 & $-141.612$ \\
7.0 & 1.73105 & 0.01438 & 65.8765 & 2.32991 & $-183.155$ \\
\end{longtable}

\begin{longtable}{L{0.0697\textwidth} L{0.0851\textwidth} L{0.1239\textwidth} L{0.1239\textwidth} L{0.1239\textwidth} L{0.1316\textwidth}}
  \hline\hline
\cellcolor{brown!30}$\hat a$ & \cellcolor{brown!30}$r_{+}/M$ & \cellcolor{brown!30}$S_{\rm BH}/M^{2}$ & \cellcolor{brown!30}$S_{\rm exact}/M^{2}$ & \cellcolor{brown!30}$S_{\mathcal{O}(\hat a)}/M^{2}$ & \cellcolor{brown!30}relative gap \\
  \hline
  \endfirsthead
  \endhead
  \hline
  \endfoot
  \hline\hline
  \caption{Exact entropy of Eq.~\eqref{eq:entropy_exact} against the first-order closed form of Eq.~\eqref{eq:entropy_pert}, for $\Lambda=-0.03$, $N=0.05$, $w=-2/3$. The last column is $|S_{\rm exact}-S_{\mathcal{O}(\hat a)}|/S_{\rm exact}$. The closed form is accurate to better than one per cent for $\hat a\lesssim1$ and degrades to about ten per cent by $\hat a=4$.}
  \label{tab:entropy}\\
  \endlastfoot
0.5 & 2.0 & 12.5664 & 12.9235 & 12.9072 & $1.26\times10^{-3}$ \\
0.5 & 4.0 & 50.2655 & 51.6730 & 51.6099 & $1.22\times10^{-3}$ \\
0.5 & 6.0 & 113.0973 & 116.5086 & 116.3434 & $1.42\times10^{-3}$ \\
1.0 & 2.0 & 12.5664 & 13.3168 & 13.2481 & $5.16\times10^{-3}$ \\
1.0 & 4.0 & 50.2655 & 53.2209 & 52.9543 & $5.01\times10^{-3}$ \\
1.0 & 6.0 & 113.0973 & 120.2909 & 119.5895 & $5.83\times10^{-3}$ \\
2.0 & 2.0 & 12.5664 & 14.2404 & 13.9299 & $2.18\times10^{-2}$ \\
2.0 & 4.0 & 50.2655 & 56.8456 & 55.6431 & $2.12\times10^{-2}$ \\
2.0 & 6.0 & 113.0973 & 129.2867 & 126.0816 & $2.48\times10^{-2}$ \\
3.0 & 2.0 & 12.5664 & 15.4168 & 14.6116 & $5.22\times10^{-2}$ \\
3.0 & 4.0 & 50.2655 & 61.4411 & 58.3319 & $5.06\times10^{-2}$ \\
3.0 & 6.0 & 113.0973 & 141.0139 & 132.5738 & $5.99\times10^{-2}$ \\
4.0 & 2.0 & 12.5664 & 16.9894 & 15.2934 & $9.98\times10^{-2}$ \\
4.0 & 4.0 & 50.2655 & 67.5419 & 61.0207 & $9.66\times10^{-2}$ \\
4.0 & 6.0 & 113.0973 & 157.2738 & 139.0659 & $1.16\times10^{-1}$ \\
\end{longtable}

\begin{longtable}{L{0.0844\textwidth} L{0.1013\textwidth} L{0.1773\textwidth} L{0.1773\textwidth} L{0.1688\textwidth}}
  \hline\hline
\cellcolor{brown!30}$\hat a$ & \cellcolor{brown!30}$r_{+}/M$ & \cellcolor{brown!30}$(dM/dS)\,M$ & \cellcolor{brown!30}$TM$ & \cellcolor{brown!30}relative gap \\
  \hline
  \endfirsthead
  \endhead
  \hline
  \endfoot
  \hline\hline
  \caption{Numerical verification of the first law with the corrected entropy of Eq.~\eqref{eq:entropy_exact}, for $\Lambda=-0.03$, $N=0.05$, $w=-2/3$. The mass derivative was taken by central differences with step $10^{-5}$. Agreement holds at the level of $10^{-12}$ to $10^{-10}$ across the admissible range. With the area law in place of Eq.~\eqref{eq:entropy_exact} the same comparison fails at the per-cent level once $\hat a\gtrsim1$.}
  \label{tab:firstlaw}\\
  \endlastfoot
0.0 & 2.0 & 0.036605637 & 0.036605637 & $4.6\times10^{-12}$ \\
0.0 & 4.0 & 0.021485917 & 0.021485917 & $1.2\times10^{-11}$ \\
0.0 & 6.0 & 0.019629110 & 0.019629110 & $1.1\times10^{-10}$ \\
1.0 & 2.0 & 0.035216052 & 0.035216052 & $5.6\times10^{-12}$ \\
1.0 & 4.0 & 0.020743978 & 0.020743978 & $6.5\times10^{-12}$ \\
1.0 & 6.0 & 0.018954242 & 0.018954242 & $5.3\times10^{-11}$ \\
2.0 & 2.0 & 0.033651753 & 0.033651753 & $9.4\times10^{-12}$ \\
2.0 & 4.0 & 0.019903913 & 0.019903913 & $2.1\times10^{-11}$ \\
2.0 & 6.0 & 0.018165362 & 0.018165362 & $4.2\times10^{-11}$ \\
3.0 & 2.0 & 0.031861704 & 0.031861704 & $1.7\times10^{-12}$ \\
3.0 & 4.0 & 0.018935311 & 0.018935311 & $3.0\times10^{-11}$ \\
3.0 & 6.0 & 0.017214998 & 0.017214998 & $2.5\times10^{-11}$ \\
4.0 & 2.0 & 0.029767576 & 0.029767576 & $3.8\times10^{-11}$ \\
4.0 & 4.0 & 0.017790287 & 0.017790287 & $4.4\times10^{-11}$ \\
4.0 & 6.0 & 0.016015984 & 0.016015984 & $3.0\times10^{-12}$ \\
5.0 & 2.0 & 0.027239006 & 0.027239006 & $4.8\times10^{-11}$ \\
5.0 & 4.0 & 0.016386122 & 0.016386122 & $7.9\times10^{-12}$ \\
5.0 & 6.0 & 0.014375234 & 0.014375234 & $4.1\times10^{-11}$ \\
\end{longtable}

\section{Joule-Thomson expansion}\label{isec6}

We can now answer the question posed in Sec.~\ref{isec1}. Treating the mass as the enthalpy of the extended phase space~\cite{Kastor:2009wy,Dolan:2011xt,Gunasekaran:2012dq}, an identification that rests on the AdS phase structure first mapped in Refs.~\cite{Hawking:1982dh,Chamblin:1999tk,Chamblin:1999hg} and on the van der Waals analogy of Ref.~\cite{Kubiznak:2012wp}, the throttling process holds $M$ fixed while the pressure varies, and the response of the temperature is measured by
\begin{equation}
\mu_{\rm JT}=\Big(\frac{\partial T}{\partial P}\Big)_{M}.
\label{eq:mu_def}
\end{equation}
The horizon radius is not held fixed: it slides with $P$ along the constraint $f(r_{+})=0$ at constant mass. Writing $\partial f/\partial P=8\pi r_{+}^{2}/3$ and using $T=f'(r_{+})/4\pi$,
\begin{equation}
\mu_{\rm JT}=\frac{\partial T}{\partial r_{+}}\Big(\frac{\partial r_{+}}{\partial P}\Big)_{M}+\frac{\partial T}{\partial P}
=\frac{f''}{4\pi}\Big(-\frac{8\pi r_{+}^{2}/3}{f'}\Big)+\frac{4r_{+}}{3},
\end{equation}
which collapses to the compact closed form
\begin{equation}
\mu_{\rm JT}=\frac{2r_{+}}{3}\left[2-\frac{r_{+}f''(r_{+})}{f'(r_{+})}\right].
\label{eq:mu_closed}
\end{equation}
Equation~\eqref{eq:mu_closed} holds for any metric of the class~\eqref{eq:metric} whose pressure enters only through the $-\Lambda r^{2}/3$ term, and we verified it symbolically as well as against central differences. Its zero locus is
\begin{equation}
r_{+}f''(r_{+})=2f'(r_{+}),
\label{eq:inversion_condition}
\end{equation}
which is the photon-sphere condition~\eqref{eq:phsphere} with $f$ replaced by $f'$. Equivalently, since $r f''-2f'=r^{3}\,d(f'/r^{2})/dr$, the inversion radius is the stationary point of $f'/r^{2}$, that is of $T/r_{+}^{2}$. In the undeformed case where $S\propto r_{+}^{2}$ this says the inversion radius extremises $T/S$. We are not aware of this correspondence being noted before, and it ties the throttling behaviour of the black hole to the null geodesic structure of the same geometry through a single derivative shift.

Applying Eq.~\eqref{eq:mu_closed} to neutral Schwarzschild-AdS, where $f=1-2M/r+8\pi Pr^{2}/3$, the horizon condition gives $f'=1/r+8\pi Pr$ and $f''=-2/r^{2}$, so
\begin{equation}
\mu_{\rm JT}^{\rm SAdS}=\frac{8r_{+}}{3}\,\frac{1+4\pi Pr_{+}^{2}}{1+8\pi Pr_{+}^{2}}>0\quad\text{for all }r_{+}.
\label{eq:mu_sads}
\end{equation}
The coefficient is positive everywhere, so the neutral Schwarzschild-AdS black hole cools under throttling at every radius and possesses no inversion point at all. This is the baseline against which the deformation has to be measured, and it is why the question of Sec.~\ref{isec1} is not idle: any inversion found below is generated by $\hat a$, not inherited from the background.

\begin{figure}[ht!]
  \centering
  \includegraphics[width=0.8\textwidth]{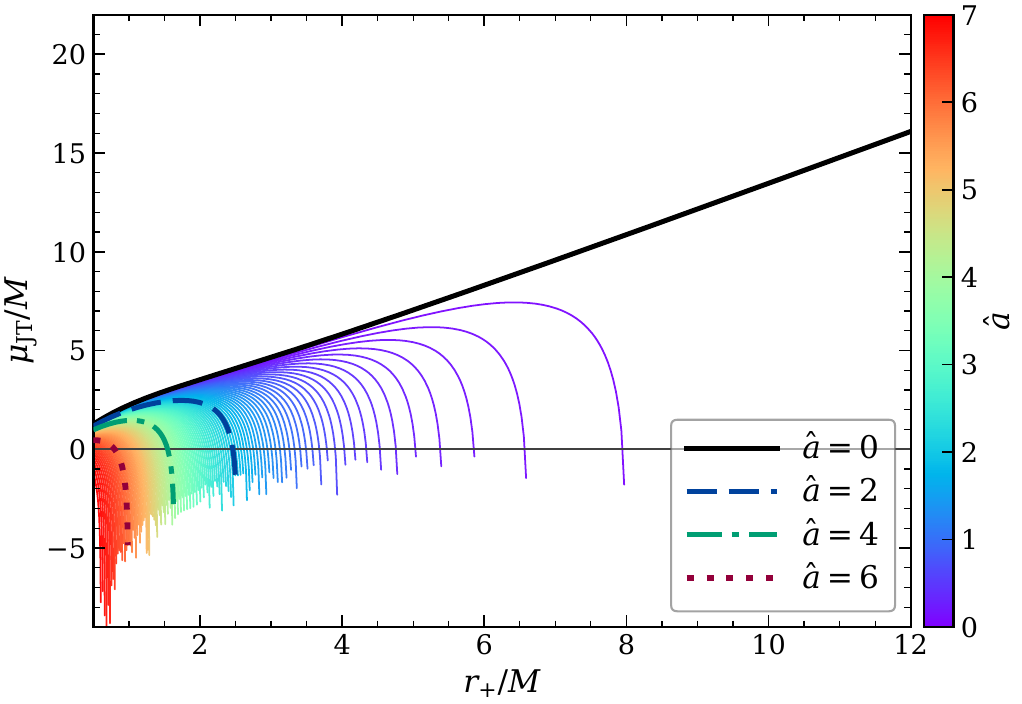}
  \caption{Joule-Thomson coefficient of Eq.~\eqref{eq:mu_closed} at $P=0.02M^{-2}$, for $N=0.05$, $w=-2/3$, with the ribbon sweeping $\hat a$ over $[0,7]$. The horizontal line marks $\mu_{\rm JT}=0$. The inset covers $r_{+}/M\in[0.8,3.2]$, where the sign change occurs and the curves cross.}
  \label{fig:muJT}
\end{figure}

Switching on the deformation changes the picture. Figure~\ref{fig:muJT} shows $\mu_{\rm JT}$ at $P=0.02M^{-2}$. At $\hat a=0$ the coefficient is positive across the whole range, as Eq.~\eqref{eq:mu_sads} requires once the quintessence tail is added. For $\hat a\gtrsim1$ a region of negative $\mu_{\rm JT}$ opens at small horizon radius, so a sufficiently small deformed black hole heats rather than cools as it expands~\cite{Yin:2018jt,Sekhmani:2023egb,Sekhmani:2026thr}, and the sign change defines the inversion radius. The mechanism is visible in Eq.~\eqref{eq:mu_closed}: the deformation contributes $+6\hat aM^{3}/\pi r^{5}$ to $f''$ and $-3\hat aM^{3}/2\pi r^{4}$ to $f'$, so it raises the ratio $r f''/f'$ steeply at small radius and drives the bracket negative.

\begin{figure}[ht!]
  \centering
  \includegraphics[width=0.8\textwidth]{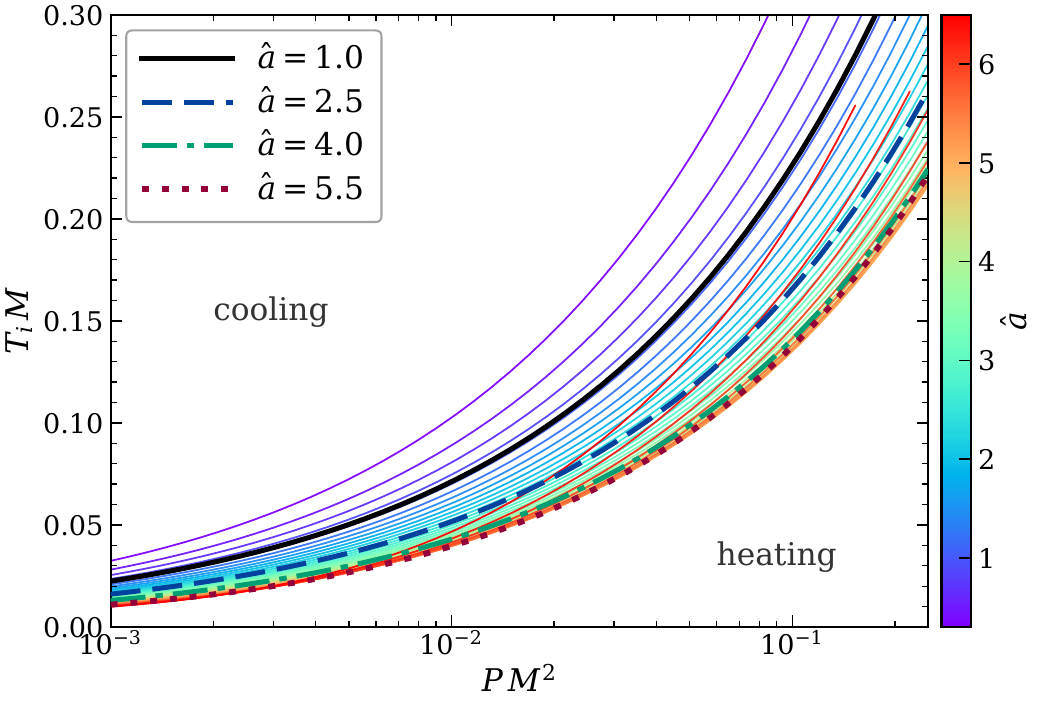}
  \caption{Inversion curves in the pressure-temperature plane for $N=0.05$, $w=-2/3$, with the ribbon sweeping $\hat a$ over $[0.3,6.5]$. The pressure axis is logarithmic. The inset covers $P M^{2}\in[10^{-3},6\times10^{-3}]$, where the curves converge at low pressure.}
  \label{fig:inversion}
\end{figure}

\begin{figure}[ht!]
  \centering
  \includegraphics[width=0.8\textwidth]{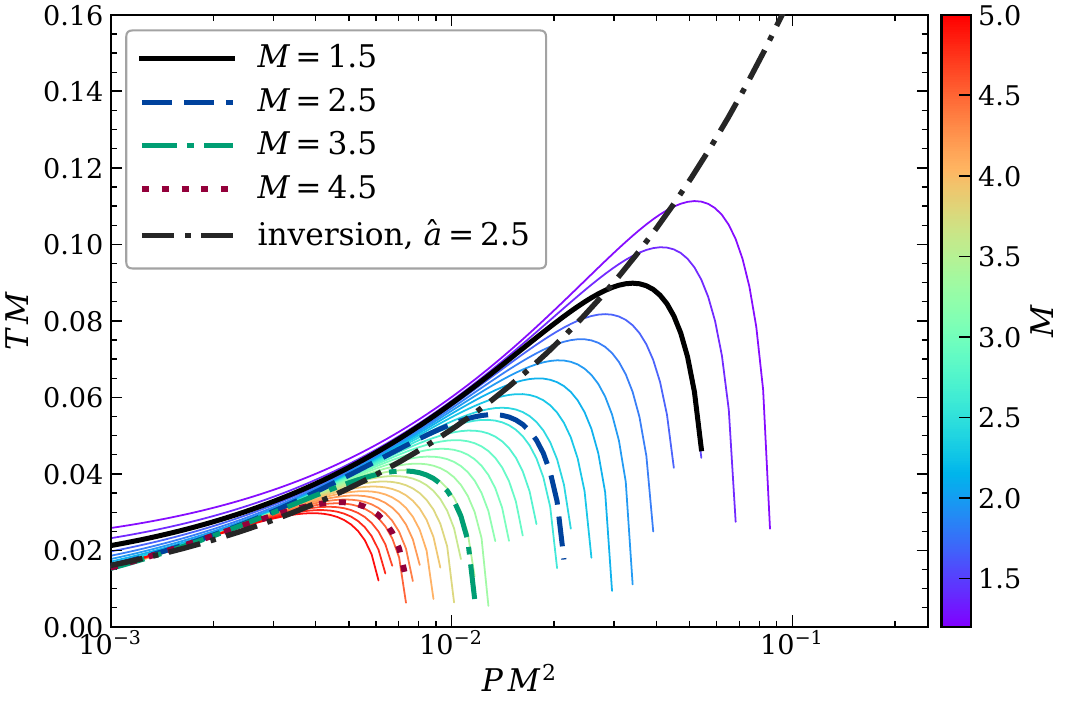}
  \caption{Isenthalpic curves at fixed mass for $\hat a=3$, $N=0.05$, $w=-2/3$, with the ribbon sweeping $M$ over $[1.2,5.0]$. The heavy dash-dotted curve is the inversion locus at $\hat a=2.5$, which passes through the maxima of the isenthalps. The inset covers $PM^{2}\in[2\times10^{-3},2\times10^{-2}]$ around those maxima.}
  \label{fig:isenthalps}
\end{figure}

Figure~\ref{fig:inversion} maps the inversion curves in the $(P,T_{i})$ plane and Table~\ref{tab:inversion} gives the numbers. Every admissible $\hat a>0$ produces an inversion curve spanning the pressure range examined. The curves are monotonically increasing in $P$, as they are for charged AdS black holes~\cite{Yin:2018jt,Sakalli:2026jt}, and they shift downward as $\hat a$ grows: at $P=0.02M^{-2}$ the inversion temperature falls from $0.10088/M$ at $\hat a=1$ to $0.05809/M$ at $\hat a=5$, while the inversion radius contracts from $3.3315M$ to $1.1842M$. The region above each curve is the cooling region and the region below is the heating region. Figure~\ref{fig:isenthalps} overlays the isenthalpic curves, along which the throttling actually proceeds; each has a maximum, and the locus of those maxima is the inversion curve, which is the standard consistency check on a Joule-Thomson analysis and which our curves pass.

So the answer to the question of Sec.~\ref{isec1} is affirmative on the first count: the $\kappa$-deformation does induce Joule-Thomson expansion in an uncharged, non-rotating AdS black hole, reproducing what the Moyal and loop-quantum-gravity corrections do. The effect itself has been catalogued across a wide range of AdS black holes, from the charged and rotating cases~\cite{Okcu:2016tgt,Okcu:2017qgo,Mo:2018rgq} through Gauss-Bonnet, Lovelock, Born-Infeld, quasitopological and massive-gravity extensions~\cite{Lan:2018nnp} to regular and deformed geometries~\cite{Pu:2019bxf,Li:2019jcd,Rajani:2020mdw,Gogoi:2026ijd,Ahmed:2025qza}, and to the noncommutative case that motivates the present question~\cite{Graca:2021ker} \cite{Wang:2024jlj,Wang:2025ycl,Wang:2024jtp,MoraisGraca:2021ife}. It is negative on the second. The Moyal and loop-quantum-gravity analyses report both a Joule-Thomson inversion and a van der Waals critical point, and quote the ratio of the minimum inversion temperature to the critical temperature as a parameter-independent number~\cite{Zhao:2014dark,Lan:2020hp,Liang:2021hp}. That ratio cannot be formed here. Criticality of this kind has been found in Gauss-Bonnet, dilatonic, quasitopological, hairy and rotating settings~\cite{Cai:2013qga,Dehghani:2014caa,Hennigar:2015esa}, and its thermodynamic geometry has been worked out for the Schwarzschild-AdS and Reissner-Nordstrom-AdS cases~\cite{Zhang:2014uoa}. Scanning the isotherms across the admissible window $\hat a<\hat a_{\max}$ and over $T\in[8\times10^{-3},3.2\times10^{-2}]M^{-1}$, we find that $P(r_{+})$ carries a single maximum and no oscillation, so there is no critical point and no first-order small-to-large transition to terminate. The absence is not a numerical artifact of a coarse grid: the admissibility bound of Eq.~\eqref{eq:amax} truncates each isotherm at the radius where the mass branch of Eq.~\eqref{eq:mass_branch} ceases to exist, and that truncation removes precisely the large-radius region in which a van der Waals loop would have to close.

The contrast is the substantive result of this section. In the Moyal case the smearing length is an independent scale, and it can generate a critical point because it survives at large horizon radius. In the $\kappa$-deformed case the amplitude is locked to $M^{3}$ and the admissibility bound shuts the solution down before the large-radius structure develops. The two deformations therefore separate at the level of the phase diagram even though they agree at the level of the throttling sign. Any claim that Joule-Thomson expansion and van der Waals criticality are two faces of one quantum-gravity correction does not survive this example.

\begin{longtable}{L{0.0518\textwidth} L{0.0888\textwidth} L{0.0962\textwidth} L{0.0888\textwidth} L{0.0962\textwidth} L{0.0888\textwidth} L{0.0962\textwidth}}
  \hline\hline
\cellcolor{brown!30}$\hat a$ & \cellcolor{brown!30}$r_{i}/M$ & \cellcolor{brown!30}$T_{i}M$ & \cellcolor{brown!30}$r_{i}/M$ & \cellcolor{brown!30}$T_{i}M$ & \cellcolor{brown!30}$r_{i}/M$ & \cellcolor{brown!30}$T_{i}M$ \\
\cellcolor{brown!30} & \multicolumn{2}{c}{\cellcolor{brown!30}$PM^{2}=0.005$} & \multicolumn{2}{c}{\cellcolor{brown!30}$PM^{2}=0.02$} & \multicolumn{2}{c}{\cellcolor{brown!30}$PM^{2}=0.10$} \\
  \hline
  \endfirsthead
  \endhead
  \hline
  \endfoot
  \hline\hline
  \caption{Inversion radii and inversion temperatures from Eq.~\eqref{eq:inversion_condition}, for $N=0.05$, $w=-2/3$. Both quantities fall as the deformation grows. At $\hat a=0$ the coefficient of Eq.~\eqref{eq:mu_sads} is positive throughout and no inversion exists, so the table starts at $\hat a=1$.}
  \label{tab:inversion}\\
  \endlastfoot
1.0 & 6.9824 & 0.05021 & 3.3315 & 0.10088 & 1.4518 & 0.22634 \\
2.0 & 5.2516 & 0.03946 & 2.4622 & 0.07965 & 1.0625 & 0.17934 \\
3.0 & 4.2243 & 0.03370 & 1.9444 & 0.06855 & 0.8305 & 0.15516 \\
4.0 & 3.4354 & 0.02992 & 1.5447 & 0.06167 & 0.6511 & 0.14085 \\
5.0 & 2.7304 & 0.02745 & 1.1842 & 0.05809 & 0.4889 & 0.13505 \\
\end{longtable}

\section{Sparsity of the radiation and the emission rate}\label{isec7}

\begin{figure}[ht!]
  \centering
  \includegraphics[width=0.72\textwidth]{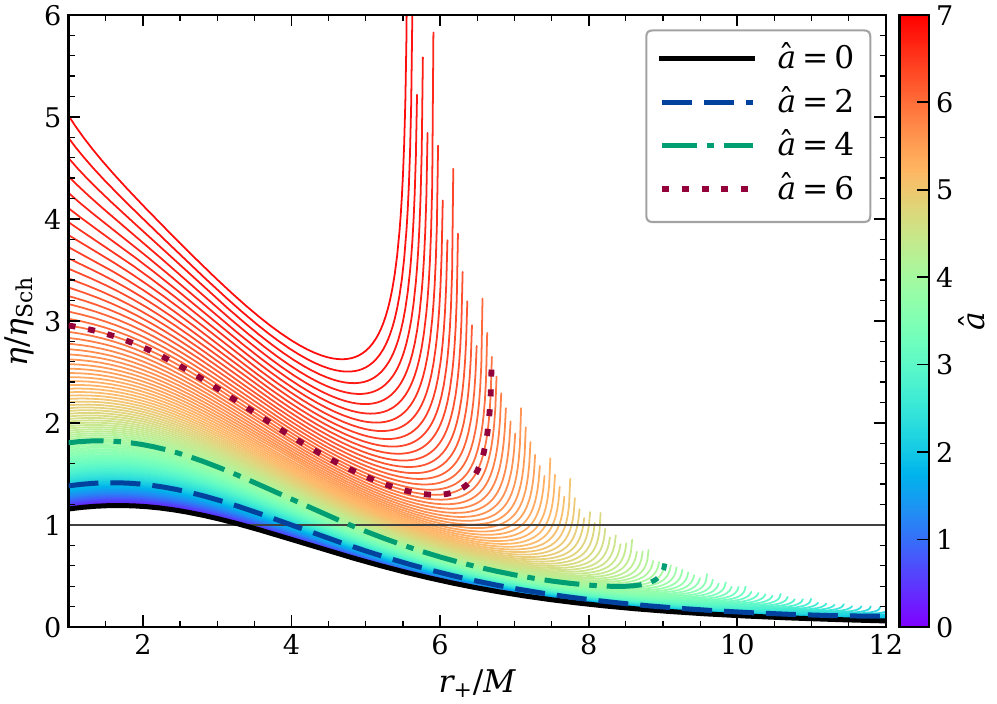}
  \caption{Sparsity of Eq.~\eqref{eq:sparsity_ratio} normalised to the Schwarzschild value $64\pi^{3}/27$, for $\Lambda=-0.03$, $N=0.05$, $w=-2/3$, with the ribbon sweeping $\hat a$ over $[0,7]$. The horizontal line marks the Schwarzschild value. The divergence is the extremal limit, where the temperature passes through zero.}
  \label{fig:sparsity}
\end{figure}

The temperature of Sec.~\ref{isec5} fixes not only how hot the hole is but how thinly its radiation arrives, and the deformation shows up in both. The thermal wavelength attached to the Hawking temperature is $\lambda_{t}=2\pi/T$, and the effective radiating area, enlarged beyond the horizon by back-scattering off the potential barrier, is $A_{\rm eff}=\tfrac{27}{4}A_{\rm BH}=27\pi r_{+}^{2}$. The dimensionless sparsity is their ratio,
\begin{equation}
\eta=\frac{\lambda_{t}^{2}}{A_{\rm eff}}=\frac{4\pi}{27\,r_{+}^{2}\,T^{2}},
\label{eq:sparsity}
\end{equation}
~\cite{Wang:2026slow,Sakalli:2020qui} measures the mean separation between successive quanta in units of the emission timescale; $\eta\gg1$ means the flux arrives as isolated quanta rather than as a stream. For Schwarzschild, $T=1/8\pi M$ and $r_{+}=2M$ give $\eta_{\rm Sch}=64\pi^{3}/27\simeq73.5$, already far above unity. Using the temperature of Eq.~\eqref{eq:temperature} with $w=-2/3$ and writing $B\equiv1+8\pi Pr_{+}^{2}-2Nr_{+}$,
\begin{equation}
\frac{\eta}{\eta_{\rm Sch}}=\left[B-\frac{\hat aM^{3}}{\pi r_{+}^{3}}\right]^{-2},
\label{eq:sparsity_ratio}
\end{equation}
the exponent being $-2$ because $\eta$ scales as $T^{-2}$. Both corrections in the bracket push in the same direction. The quintessence contributes $-2Nr_{+}$ and the deformation $-\hat aM^{3}/\pi r_{+}^{3}$, so each lowers the temperature and raises the sparsity, and a deformed hole radiates more thinly than its undeformed counterpart at the same horizon radius. Figure~\ref{fig:sparsity} shows the ratio. It rises steeply as $\hat a$ grows and diverges where the temperature passes through zero, which is the extremal limit approached along the small-$r_{+}$ branch; near that point the emission becomes arbitrarily sparse and the semiclassical description of a thermal flux loses its meaning.

\begin{figure}[ht!]
  \centering
  \includegraphics[width=0.72\textwidth]{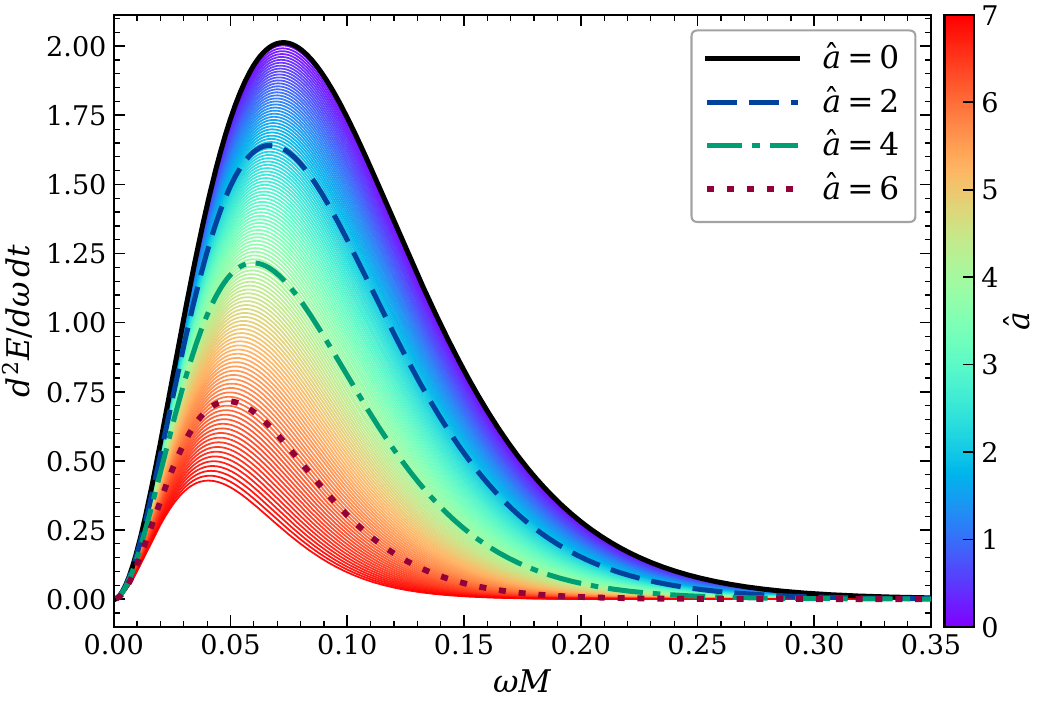}
  \caption{Energy emission rate of Eq.~\eqref{eq:emission} at $r_{+}=3M$, for $\Lambda=-0.03$, $N=0.05$, $w=-2/3$, with the ribbon sweeping $\hat a$ over $[0,7]$. The shadow radius is evaluated for a static observer at $r_{0}=50M$.}
  \label{fig:emission}
\end{figure}

The energy carried away follows from the same temperature together with the shadow of Sec.~\ref{isec3}. In the high-frequency regime the absorption cross section approaches the geometric value set by the shadow, $\sigma_{\rm lim}\simeq\pi R_{\rm sh}^{2}$, ~\cite{Nozari:2024ncq,Sood:2024sha}, so the emission rate per unit frequency is
\begin{equation}
\frac{d^{2}E}{d\omega\,dt}=\frac{2\pi^{2}\sigma_{\rm lim}\,\omega^{3}}{e^{\omega/T}-1}
=\frac{2\pi^{3}R_{\rm sh}^{2}\,\omega^{3}}{e^{\omega/T}-1},
\label{eq:emission}
\end{equation}
with $R_{\rm sh}=r_{\rm ph}\sqrt{f(r_{0})/f(r_{\rm ph})}$ and $T$ from Eq.~\eqref{eq:temperature}. The deformation enters twice and with opposite signs in principle, through the prefactor $R_{\rm sh}^{2}$ and through the exponent. In practice the exponent wins by a wide margin. Figure~\ref{fig:emission} shows the spectrum at $r_{+}=3M$. The peak drops from $1.96$ at $\hat a=0$ to $0.51$ at $\hat a=6$, a reduction by a factor close to four, and it moves to lower frequency, from $\omega M=0.070$ to $\omega M=0.043$. The shadow contracts by only about a sixteenth over the same range, as Table~\ref{tab:null} records, so essentially all of the suppression traces to the temperature drop of Eq.~\eqref{eq:temperature} acting inside the Planck factor. A deformed black hole therefore evaporates more slowly and more coldly, and the two effects reinforce rather than cancel.

\section{Conclusion}\label{isec8}

We have studied the $\kappa$-deformed Schwarzschild-AdS black hole immersed in a Kiselev quintessence field, following the geometry from its curvature invariants through its null and timelike geodesics into its thermodynamics and its behaviour under throttling. The organising feature throughout has been the way the deformation enters, as a term $\hat aM^{3}/2\pi r^{3}$ whose amplitude is tied to the mass rather than standing as an independent hair. That single structural fact generates every result reported here.

It first constrains where the solution exists. Because the horizon condition becomes a cubic in the mass, the branch connected to the undeformed solution survives only for $\hat a<64\pi/27G^{2}$, which in the Schwarzschild sector is the pure number $7.4467$ and which we located independently as the point where the inner and outer horizons merge at $4M/3$. Since $G$ grows with horizon radius once the cosmological term takes over, the bound also caps the size of the hole, and it is this truncation that shapes the phase diagram later on. Within that window the deformation opens an inner Cauchy horizon in a geometry carrying neither charge nor rotation, and it narrows the region between the horizons by more than a factor of two across the admissible range. What it does not do is remove the central singularity. The Kretschmann invariant behaves as $46\hat a^{2}M^{6}/\pi^{2}r^{10}$ near the origin, a stronger divergence than the Schwarzschild $48M^{2}/r^{6}$, and the bounce reported for $\kappa$-Minkowski cosmology has no counterpart here. This is the generic fate of a power-law addition to the lapse, and it is worth stating clearly because minimal-length corrections are often assumed to regularise what they modify. The effective source is better behaved than the singularity suggests. The tangential null energy combination reduces to $(5\hat aM^{3}/r^{5}-9\pi Nw(w+1)r^{-3w-3})/16\pi^{2}$, in which the mass and the cosmological constant cancel identically, so it is non-negative for every admissible $\hat a$ and for quintessence proper, vanishing only in the Schwarzschild-anti-de Sitter limit.

On the optical side the deformation contracts the photon sphere from $3.2668M$ to $2.8664M$ and the critical impact parameter from $5.6768M$ to $5.3374M$ over the admissible range, with the angular radius seen by a static observer at $50M$ falling from $33.36^{\circ}$ to $31.13^{\circ}$. The Lyapunov exponent decreases while the orbital frequency at the photon sphere increases, so at the level of the eikonal estimate the deformation raises the ringdown frequency and lengthens the damping time together. We have been careful to treat the shadow correctly in a background that is not asymptotically flat, quoting angular radii with the observer position attached, and to note that at $w=-1/3$ with vanishing cosmological constant the solid angle deficit $f(\infty)=1-N$ enters the shadow through a square root that reverses how it scales with the quintessence normalisation. For massive particles the innermost stable circular orbit contracts from $4.6072M$ to $4.2021M$, a shift of under a tenth, so twin-peak quasiperiodic oscillations constrain $\hat a$ far less tightly than the shadow does. The quintessence tail also produces a second, outer unstable photon orbit near $r\simeq2/N$, and circular timelike orbits are confined between the two.

The thermodynamic consequence is the sharpest of the results. Since the temperature is fixed by the surface gravity and the deformation amplitude varies with the mass, the Bekenstein-Hawking area law and the first law cannot both hold. Requiring $dM=T\,dS$ selects the entropy $S=\int2\pi r^{3}dr/(r^{2}-3\hat aM^{2}/4\pi)$, which we verified against the first law to between $10^{-12}$ and $10^{-10}$ across six deformation strengths and three horizon radii, and which exceeds the area law by a factor $1.34$ at $\hat a=4$ and $r_{+}=3M$. Expanding to first order gives a closed form that is accurate to better than a per cent for $\hat a\lesssim1$, and in the Schwarzschild sector it reduces to the pure rescaling $\pi r_{+}^{2}(1+3\hat a/16\pi)$ rather than to the logarithmic correction that minimal-length arguments usually produce. We have also set out the alternative prescription in which the combination $\hat aM^{3}/2\pi$ is held fixed as an independent hair, under which the area law survives intact, and explained why the construction of the solution does not license it.

For the throttling problem we obtained the Joule-Thomson coefficient in closed form, $\mu_{\rm JT}=(2r_{+}/3)[2-r_{+}f''/f']$, valid for any metric of this class. Its zero locus, $r_{+}f''=2f'$, is the photon-sphere condition $rf'=2f$ shifted by one derivative, so the inversion radius is the stationary point of $T/r_{+}^{2}$. Neutral Schwarzschild-AdS gives $\mu_{\rm JT}=(8r/3)(1+4\pi Pr^{2})/(1+8\pi Pr^{2})$, positive at every radius and therefore free of any inversion, which makes the baseline unambiguous. Switching on the deformation opens a heating region at small radius and produces an inversion curve for every admissible $\hat a$, so the $\kappa$-deformation reproduces the Joule-Thomson behaviour of the Moyal and loop-quantum-gravity corrections. It does not reproduce their phase structure. Across the admissible window the isotherms carry a single maximum and no van der Waals oscillation, so there is no critical point, and the ratio of minimum inversion temperature to critical temperature that those analyses quote as parameter-independent simply cannot be formed. The admissibility bound removes the large-radius region in which such a loop would have to close. Joule-Thomson expansion and van der Waals criticality are therefore not two faces of a single quantum-gravity correction, and this geometry is the counterexample.
Two further results attach to the same structure. The consistency of the thermodynamics can be carried either by the entropy or by the mass, and the two are the same statement: keeping the area law and rescaling the enthalpy through $d\mathcal{M}=W\,dM$ with $W=1-3\hat aM^{2}/4\pi r_{+}^{2}$ reproduces the corrected entropy exactly, since $dS/dr_{+}=2\pi r_{+}/W$. Working in the rescaled frame is the more convenient of the two, because the conjugates come out clean: the temperature returns the surface-gravity value, the volume returns the geometric $4\pi r_{+}^{3}/3$, the quintessence potential is $-r_{+}^{-3w}/2$, and the Smarr relation reads $\mathcal{M}=2(TS-PV)+(3w+1)N\Phi_{N}$ with no deformation term, since $\hat a$ carries no length dimension and enters only through $W$. And the radiation itself is measurably altered. The sparsity rises as $[B-\hat aM^{3}/\pi r_{+}^{3}]^{-2}$, so a deformed hole emits its quanta more thinly than an undeformed one at the same horizon radius, and the peak of the emission spectrum at $r_{+}=3M$ drops by a factor close to four between $\hat a=0$ and $\hat a=6$ while shifting from $\omega M=0.070$ to $\omega M=0.043$. Since the shadow contracts by only a sixteenth over that range, the suppression is almost entirely the temperature acting inside the Planck factor rather than the geometric cross section.

Four calculations follow directly from what is established here. The eikonal frequencies of Table~\ref{tab:qnm} are photon-sphere data expressed in frequency units, and a proper treatment of scalar, electromagnetic and Dirac perturbations with reflective boundary conditions at the AdS boundary, computed by a spectral or continued-fraction method rather than by the eikonal formula, would test how far that shortcut can be trusted in this geometry and would settle whether the deformation destabilises the Cauchy horizon it creates. The inner horizon itself invites a strong cosmic censorship analysis, since the deformation supplies a Cauchy horizon without charge or rotation and the usual charge-driven mechanism for violating the censorship conjecture is absent. On the thermodynamic side, the corrected entropy and the rescaling factor $W$ should both be fed back into a Euclidean action computation, to check whether that entropy emerges as a Wald entropy or requires a boundary contribution. The same factor governs the evaporation history, which we have not worked out: a sparsity that diverges at extremality means the semiclassical description of a thermal flux fails before the hole disappears. Finally, the contraction of the shadow by roughly seven per cent across the admissible range is within reach of next-generation very-long-baseline interferometry, and folding the bound of Eq.~\eqref{eq:amax} together with existing shadow-size measurements would convert it into a direct constraint on the $\kappa$-deformation scale.

\begin{acknowledgments}
F.A.\ thanks the Inter-University Centre for Astronomy and Astrophysics, Pune, for a visiting associateship. \.{I}.S.\ is grateful to Eastern Mediterranean University, T\"UB\.ITAK, ANKOS and SCOAP3 for their support, and acknowledges the networking support of COST Actions CA18108 (``Quantum gravity phenomenology in the multi-messenger approach''), CA22113 (``Fundamental challenges in theoretical physics''), CA21106 (``COSMIC WISPers in the Dark Universe''), CA23130 (``Bridging high and low energies in search of quantum gravity''), CA23115 (``Relativistic Quantum Information''), and CA21136 (``Addressing observational tensions in cosmology with systematics and fundamental physics''). We acknowledge the INSPIRE-HEP database for bibliographic resources.
\end{acknowledgments}

\section*{Data Availability Statement}

No new observational data were generated in this study. All results reported here follow from the analytical expressions given explicitly in Secs.~\ref{isec2} through~\ref{isec7}, together with the parameter values stated in the figure and table captions. The computational worksheets used to produce Figs.~\ref{fig:lapse} through~\ref{fig:emission} and Tables~\ref{tab:limits} through~\ref{tab:inversion} are available from the corresponding author upon reasonable request.

\IfFileExists{apsrev4-2.bst}{%
  \bibliographystyle{apsrev4-2}%
}{%
  \IfFileExists{apsrev4-1.bst}{%
    \bibliographystyle{apsrev4-1}%
  }{%
    \bibliographystyle{unsrt}%
  }%
}
\bibliography{finalref}

\end{document}